\documentclass[runningheads]{llncs}

\usepackage{graphicx}
\usepackage{epsfig,epsf}
\usepackage{epstopdf}
\usepackage{graphics}
\usepackage{array}
\usepackage{amsmath}
\usepackage{amssymb}
\usepackage{comment}
\usepackage{tikz}
\usepackage{fullpage}
\usepackage{multirow}
\usepackage{enumitem}
\usepackage{bm}

\makeatletter
\renewcommand\subsubsection{\@startsection{subsubsection}{3}{\z@}%
	{-18\p@ \@plus -4\p@ \@minus -4\p@}%
	{0.5em \@plus 0.22em \@minus 0.1em}%
	{\normalfont\normalsize\bfseries\boldmath}}
\makeatother
\usepackage{amsmath,amssymb}
\usepackage{mathtools}
\usepackage{setspace}
\usepackage{array}
\usepackage{subcaption}
\usepackage{microtype}
\usepackage{graphics}
\usepackage{cite}

\usepackage[bookmarks=false]{hyperref} 
\usepackage{algorithm}
\usepackage{algpseudocode}
\usepackage{tikz}
\usepackage{adjustbox}

\newtheorem{mclaim}[theorem]{Claim}

\newcounter{claimcounter}
\numberwithin{claimcounter}{theorem}
\newenvironment{myclaim}{\stepcounter{claimcounter}{\noindent\textbf{Claim \theclaimcounter.}}}{}
\usepackage{xcolor}

\begin{document}
	\title{On Convexity in Split graphs: Complexity of Steiner tree and Domination
		%P versus NPC: Minimum Steiner trees in convex split graphs
		\thanks{This work is partially supported by the DST-ECRA Project—
			ECR/2017/001442.}\protect\footnote{A preliminary version of this paper appeared in the proceedings of $8^{th}$ International conference, CALDAM 2022, Lecture Notes in Computer Science, vol. 13179, pp. 128-139, 2022}}
		%Convexity in split graphs - A framework for Steiner Tree and Domination problems
	\author{A Mohanapriya \inst{1}\and
		P Renjith \inst{2}\and
		N Sadagopan \inst{1}}
	\institute{Indian Institute of Information Technology, Design and Manufacturing, Kancheepuram, Chennai. \and
		National Institute of Technology, Calicut
		\\\email{coe19d003@iiitdm.ac.in, renjith@nitc.ac.in, sadagopan@iiitdm.ac.in}}
	\maketitle
	\begin{abstract}
		Given a graph $G$ with a terminal set $R \subseteq V(G)$, the Steiner tree problem (STREE) asks for a set $S\subseteq  V(G) \setminus R$ such that the graph induced on $S\cup R$ is connected.  A split graph is a graph which can be partitioned into a clique and an independent set.  It is known that STREE is NP-complete on split graphs \cite{white1985steiner}.  To strengthen this result, we introduce convex ordering on one of the partitions (clique or independent set), and prove that STREE is polynomial-time solvable for tree-convex split graphs with convexity on clique ($K$), whereas STREE is NP-complete on tree-convex split graphs with convexity on independent set ($I$).  We further strengthen our NP-complete result by establishing a dichotomy which says that for unary-tree-convex split graphs (path-convex split graphs), STREE is polynomial-time solvable, and NP-complete for binary-tree-convex split graphs (comb-convex split graphs).  We also show that STREE is polynomial-time solvable for triad-convex split graphs with convexity on $I$, and circular-convex split graphs.  Further, we show that STREE can be used as a framework for the dominating set problem (DS) on split graphs, and hence the classical complexity (P vs NPC) of STREE and DS is the same for all these subclasses of split graphs.  Furthermore, it is important to highlight that in \cite{CHLEBIK20081264}, it is incorrectly claimed that the problem of finding a minimum dominating set on split graphs cannot be approximated within $(1-\epsilon)\ln |V(G)|$ in polynomial-time for any $\epsilon >0$ unless NP $\subseteq$ DTIME $n^{O(\log \log n)}$. When the input is restricted to split graphs, we show that the minimum dominating set problem has $2-\frac{1}{|I|}$-approximation algorithm that runs in polynomial time.  Finally, from the parameterized perspective with solution size being the parameter, we show that the Steiner tree problem on split graphs is $W[2]$-hard, whereas when the parameter is treewidth and the solution size, we show that the problem is fixed-parameter tractable, and if the parameter is the solution size and the maximum degree of $I$ ($d$), then we show that the Steiner tree problem on split graphs has a kernel of size at most $(2d-1)k^{d-1}+k,~k=|S|$.
		\\\textbf{Keywords:} Steiner tree, Domination, Split graphs, Tree-convex, Circular-convex split graphs, Approximation algorithms, Parameterized complexity.
	\end{abstract}
	\section{Introduction}
	The classical complexity of the Steiner tree problem (STREE), the dominating set problem (DS), and their variants for different classes of graphs have been well studied.  Given a graph $G$ with a terminal set $R \subseteq V(G)$, STREE asks for a set $S\subseteq  V(G) \setminus R$ such that the graph induced on $S\cup R$ is connected.  In the literature, the set $S$ is referred to as the Steiner set.  The objective is to minimize the number of vertices in $S$.  STREE is NP-complete for general graphs, chordal bipartite graphs \cite{muller1987np}, and split graphs \cite{white1985steiner} whose vertex set can be partitioned into a clique and an independent set.   It is polynomial-time solvable in strongly chordal graphs \cite{white1985steiner}, series-parallel graphs \cite{wald1983steiner}, outerplanar graphs \cite{wald1982steiner}, interval graphs \cite{mohanapriya2021steiner} and for graphs with fixed treewidth \cite{chimani2012improved}.  The only known subclass of split graphs where STREE is polynomial-time solvable is the class of threshold graphs.  Interestingly the results of \cite{renjith2020steiner} strengthen the result of \cite{white1985steiner} by providing a dichotomy result which says that STREE is polynomial-time solvable in $K_{1,4}$-free split graphs, whereas in $K_{1,5}$-free split graphs, STREE is NP-complete.  In this paper, we focus on new subclasses of split graphs and study the tractability versus intractability status (P vs NPC) of STREE in those subclasses of split graphs.\\
	It is important to highlight that many problems that are NP-complete on bipartite graphs become polynomial-time solvable when a linear ordering is imposed on one of the partitions.  Such graphs are known as convex bipartite graphs in the literature \cite{damaschke1990domination, panda2021dominating, chen2016complexity}.   For example, DS is NP-complete on bipartite graphs,  whereas it is polynomial-time solvable in convex bipartite graphs \cite{damaschke1990domination}.  A bipartite graph $G=(X,Y)$ is said to be tree-convex if there is a tree (imaginary tree) on $X$ such that the neighborhood of each $y$ in $Y$ is a subtree in $X$.   Apart from linear ordering (path-convex ordering), tree-convex ordering, comb-convex ordering, star-convex ordering, triad-convex ordering, and circular-convex ordering on bipartite graphs have been considered in the literature \cite{pandey2019domination, jiang2011tractable, jiang2011two}.    Further, the convex ordering on bipartite graphs yielded many interesting algorithmic results for STREE, DS, Hamiltonicity, and its variants \cite{mohanapriya2021steiner,pandey2019domination,chen2016complexity}.   Similarly, the feedback vertex set problem (FVS) is NP-complete on star-convex bipartite graphs, and comb-convex bipartite graphs, whereas it is polynomial-time solvable on convex bipartite graphs \cite{chen2016complexity}.  Thus, the convex ordering on bipartite graphs reinforces the borderline separating P-versus-NPC instances of many classical combinatorial problems.
	\\Imposing the property convexity on bipartite graphs is a promising direction for further research because many problems that are NP-complete on bipartite graphs become polynomial-time solvable on convex bipartite graphs.  Further, some of the NP-hard reductions restricted to bipartite graphs can be reinforced further by introducing convex properties such as star, comb, tree, etc.,  For example, Hamiltonian cycle and Hamiltonian path are NP-hard on star-convex bipartite graphs \cite{chen2016complexity}.  While convexity in bipartite graphs seems to be a promising direction in strengthening the existing classical hardness result or in discovering a polynomial-time algorithm,  we wish to investigate this line of research for STREE and DS problems restricted to split graphs.
	\\
	Since the tractability versus intractability status of many combinatorial problems on bipartite graphs (graphs with two partitions satisfying some structural properties) can be investigated with the help of convex ordering on bipartite graphs, it is natural to explore this line of study on graphs having two partitions satisfying some structural properties.  A natural choice after bipartite graphs is the class of split graphs.  We wish to extend this line of study to split graphs by considering convex ordering with respect to the clique part and independent set part.   To the best of our knowledge, this paper makes the first attempt in introducing convex properties on split graphs for STREE and DS.  We believe that our results shall strengthen the result of \cite{white1985steiner}, and also we discover a dichotomy similar to \cite{renjith2020steiner}.  As part of this paper, we consider the following convex properties; path-convex, star-convex, comb-convex, tree-convex, and circular-convex split graphs.  Henceforth, we refer to split graphs satisfying some convex properties (path, star, comb, triad, tree, and circular) as convex split graphs.\\
	Recently in \cite{mohanapriya2021steiner}, a framework for STREE and DS was developed, and as per \cite{renjith2020steiner}, the classical complexity of STREE is the same as the classical complexity of DS for split graphs.  We attempt a similar framework for STREE and DS, and its variants are restricted to convex split graphs.
	\\For tree-convex and its subclasses, and circular-convex split graphs, the computational complexity of the following graph problems is studied in this paper.
	\begin{enumerate}
		\item The Steiner tree problem (STREE).
		\\\emph{Instance:} A graph $G$, a terminal set $R\subseteq V(G)$, and a positive integer $k$.
		\\\emph{Question:} Does there exist a set $S\subseteq V(G)\setminus R$ such that $|S|\leq k$, and $G[S\cup R]$ is connected ?
		\item The Dominating set problem (DS).
		\\\emph{Instance:} A graph $G$, and a positive integer $k$.
		\\\emph{Question:} Does $G$ admit a dominating set of size at most $k$ ?
		\item The Connected Dominating set problem (CDS).
		\\\emph{Instance:} A graph $G$, and a positive integer $k$.
		\\\emph{Question:} Does $G$ admit a connected dominating set of size at most $k$ ?
		\item The Total Dominating set problem (TDS).
		\\\emph{Instance:} A graph $G$, and a positive integer $k$.
		\\\emph{Question:} Does $G$ admit a total dominating set of size at most $k$ ?
	\end{enumerate}
	\begin{figure}[htbp] 
		\begin{center}
			\includegraphics[scale=1]{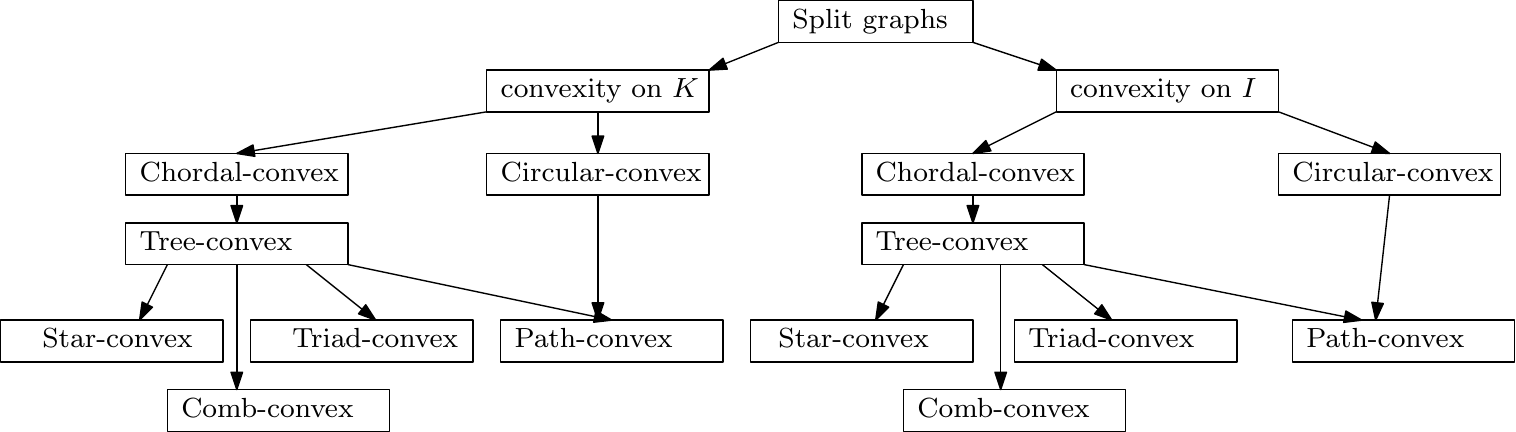}
			\caption{The Hierarchical relationship among subclasses of convex split graphs}\label{split}
		\end{center}
	\end{figure}
	\noindent Figure \ref{split} illustrates the hierarchical relationship on various convex split graphs.  An interesting theoretical question is
	\\
	\\\emph{-What is the boundary between the tractability and intractability of STREE in split graphs when convex ordering is imposed on one of the partitions ?}\\
	\\In this paper, we answer this question by imposing a convex ordering on clique or independent set.  In particular,  we show that STREE is polynomial-time solvable for tree-convex split graphs with convexity on $K$, and is NP-complete for star-convex and comb-convex split graphs, and thus for tree-convex split graphs with convexity on $I$.  Further, we investigate path, triad, and circular-convex properties, and show that STREE is polynomial-time solvable for triad, path-convex split graphs with convexity on $I$, circular-convex split graphs with convexity on $I$, and circular-convex split graphs with convexity on $K$.  We then ask
	\\
	\\\emph{-For which convex property on split graphs with convexity on $K$,  STREE is intractable?}\\
	\\In this paper, we show that if the convex property is chordality, then STREE is NP-complete for chordal-convex split graphs with convexity on $K$.\\
	\\
	To deal with computationally intractable problems, the practical approach is to use approximation algorithms or parameterized algorithms.  Algorithms that output near-optimal solutions in polynomial time are precisely the class of approximation algorithms.  It is known \cite{cormen2009introduction}, that DS has an approximation algorithm with approximation ratio $(1+\ln n)$ on general graphs.  On the negative side, DS does not admit $(1-\epsilon)\ln n $ on general graphs, for any $\epsilon>0$ unless NP $\subseteq$ DTIME ($n^{O(\log \log n)}$) \cite{CHLEBIK20081264}.  In this paper, restricted to split graphs, we prove that DS exhibits $2-\frac{1}{|I|}$-approximation algorithm.  
	\\
	For decision problems with input size $n$, and a parameter $k$ (which can be a tuple of parameters), the goal of parameterized algorithms is to obtain an algorithm with runtime $f(k)n^{O(1)}$, where $f$ is a function of $k$ and independent of $n$.  Problems having such algorithms are Fixed-Parameter Tractable (FPT).  There is a hierarchy of intractable parameterized problem classes above FPT \cite{raman2008short}, they are:
	$$\mbox{FPT}\subseteq \mbox{M[1]}\subseteq \mbox{W[1]}\subseteq \mbox{M[2]}\subseteq \mbox{W[2]}\subseteq\ldots \subseteq \mbox{W[P]}\subseteq \mbox{XP}.$$
	In \cite{dreyfus1971steiner} it is shown that STREE in general graphs is in FPT if the parameter is the size of the terminal set.
	It is known \cite{dom2009incompressibility} that STREE in general graphs with parameter $|S|$ (solution size) is W[2]-hard.  We strengthen the result of \cite{dom2009incompressibility} by proving that the Steiner tree problem on split graphs is still W[2]-hard with the parameter being the solution size.
	Further, the parameterized Steiner tree problem is in FPT, when parameters are\\
	(i) the solution size and the treewidth,\\
	(ii) the solution size and the maximum degree of $I$.
	\\We reiterate that our FPT results for STREE are true for DS as well, restricted to split graphs.
	\\\\
	\noindent This paper is structured as follows: In Section \ref{ccstree}, we analyze the classical complexity of STREE on convex split graphs and present dichotomy results for convex split graphs with convexity on $I$ as well as for convex split graphs with convexity on $K$.  We also identify polynomial-time solvable instances and FPT instances of STREE on star-convex split graphs with convexity on $I$ which we present in Section \ref{ssi}, and we also prove that the Steiner tree problem with the parameter being solution size and backbone path length on comb-convex split graphs is in XP in Section \ref{scs}.  We then present results on the dominating set problem and its variants on convex split graphs in Section \ref{DS}.  In Section \ref{pc}, we present parameterized hardness of STREE on split graphs, and we also identified parameters for which parameterized version of STREE on split graphs becomes fixed-parameter tractable.  Further, we present $2-\frac{1}{I}$-approximation algorithm for domination on split graphs in Section \ref{approx}.
	\\\\
	\textbf{Graph preliminaries:}
	In this paper, we consider connected, undirected, unweighted, and simple graphs.  For a graph $G$, $V(G)$ denotes the vertex set, and $E(G)$ represents the edge set.  For a set $S \subseteq V(G)$, $G[S]$ denotes the subgraph of $G$ induced on the vertex set $S$.  The open neighborhood of a vertex $v$ in $G$ is $N_G(v)=\{u ~\vert ~\{u,v\}\in E(G)\}$ and the closed neighborhood of $v$ in $G$ is $N_G[v]=\{v\} \cup N_G(v)$.  The degree of vertex $v$ in $G$ is $d_G(v)= |N_G(v)|$.   A split graph $G$ is a graph in which $V(G)$ is partitioned into two sets; a clique $K$ and an independent set $I$. In a split graph, for each vertex $u$ in $K$, $N^I_G(u)=N_G(u) \cap I$, $d^I_G(u)=\vert N^I_G(u) \vert$, and for each vertex $v$ in $I$, $N^K_G(v)=N_G(v) \cap K$, $d^K_G(v)=\vert N^K_G(v) \vert$.   For each vertex $u$ in $K$, $N^I_G[u]=(N_G(u) \cap I)\cup \{u\}$, and for each vertex $v$ in $I$, $N^K_G[v]=(N_G(v) \cap K) \cup \{v\}$.   For a split graph $G$, $\Delta^I_G=\mbox{max}\{d^I_G(u)\}, u \in K$ and $\Delta^K_G=\mbox{max}\{d^K_G(v)\}, v \in I$.  %The notation $K_{i,j}$ refers to a complete bipartite graph with $i$ vertices in one partition and $j$ vertices in the other.
	For a set $S$, $G-S$ denotes the graph induced on $V(G)\setminus S$.  For $A=\{x_1,\ldots,x_p\}$, $\max(x_1,\ldots,x_p)$ is $x_p$; the vertex having largest index.  
	\\
	A tree is a connected acyclic graph.  A path is a tree $T$ with $V(T)=\{v_1,\ldots,v_n\},~n\geq 1$ and $E(T)=\{\{v_i,v_{i+1},~1\leq i \leq n-1\}\}$.  A cycle is a graph $C$ with $V(C)=\{v_1,\ldots,v_n\},~n\geq 3$ and $E(C)=\{\{v_i,v_{i+1},~1\leq i \leq n-1\}\}\cup \{\{v_n,v_1\}\}$.  We consider three special kinds of trees, namely, star, comb, and triad.  A star is a tree $T$ with $V(T)=\{v_1,\ldots,v_n\},~n\geq 2$ and $E(T)=\{\{v_1,v_i\}\mid 2\leq i \leq n\}$.  The root of $T$ is $v_1$ and $v_2,\ldots,v_n$ are the pendant vertices in $T$.  A comb is a tree $T$ with $V(T)=\{v_1,\ldots,v_{2n}\}$ and $E(T)=\{\{v_i,v_{n+i}\}\mid 1\leq i \leq n\}\cup \{\{v_i,v_{i+1}\}\mid 1\leq i < n\}$.  The path on $\{v_1,v_2,\ldots, v_n\},~n\geq 1$ is the backbone of the comb, and $\{v_{n+1},v_{n+2},\ldots,v_{2n}\},~n\geq 1$ are the teeth of the comb.  A triad is a tree $T$ with $V(T)=\{u,v_1,\ldots,v_p,w_1,\ldots,w_q,x_1,\ldots,x_r\},~p\geq 2,~q\geq 2,~r\geq 2$ and $E(T)=\{\{u,v_1\},\{u,w_1\},\{u,x_1\}\} \cup \cup \{\{v_i,v_{i+1}\}\mid 1\leq i \leq p-1\}\cup \{\{w_i,w_{i+1}\}\mid 1\leq i \leq q-1\}\cup \{\{x_i,x_{i+1}\}\mid 1\leq i \leq r-1\}$.
	\begin{figure}[H]
		\begin{center}
			\includegraphics[scale=0.8]{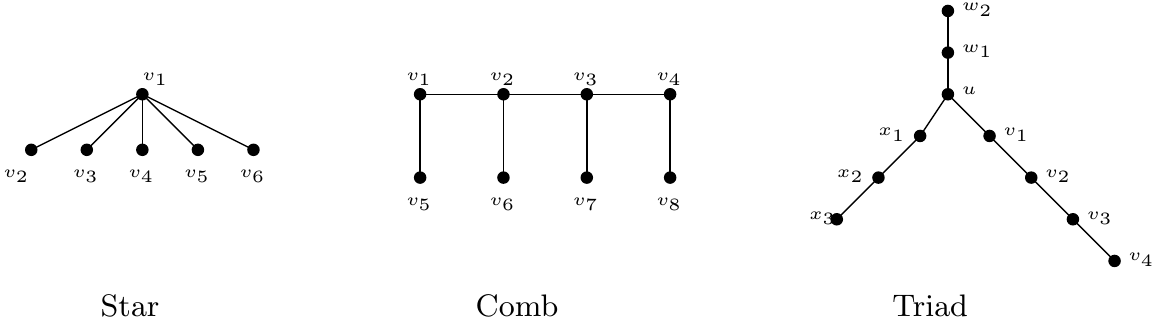}
			\caption{An example; Star, Comb, and Triad}
			\label{fig0}
		\end{center}
	\end{figure}
	%\begin{definition}
	%	A split graph $G=(K \cup I,E)$ is called {\em tree-convex with convexity on $K$} if there is an associated tree $T=(K,F)$ in $K$ such that, for each vertex $v\in I$ its neighborhood $N_G(v)$ induces a subtree of $T$.
	%\end{definition}
	%\begin{definition}
	%	A split graph $G=(K\cup I,E)$ is called {\em tree-convex with convexity on $I$} if there is an associated tree $T=(I,F)$ in $I$ such that, for each vertex $v\in K$, its neighborhood $N^I_G(v)$ induces a subtree of $T$.
	%\end{definition}
	%\noindent When $T$ is a star (caterpillar, path, triad, respectively), $G$ is called {\em star-convex} ({\em caterpillar-convex}, {\em path-convex}, {\em triad-convex, respectively}).
	%\begin{definition}
	%	A split graph $G=(K\cup I,E)$ is called {\em circular-convex with convexity on $K$} if there is an associated cycle $C=(K,F)$ in $K$ such that, for each vertex $v \in I$, its neighborhood $N_G(v)$ consists of consecutive vertices in $C$.
	%\end{definition}
	%\begin{definition}
	%	A split graph $G=(K\cup I,E)$ is called {\em circular-convex with convexity on $I$} if there is an associated cycle $C=(I,F)$ in $I$ such that, for each vertex $v\in K$, its neighborhood $N^I_G(v)$ consists of consecutive vertices in $C$.
	%\end{definition}
	\begin{definition}
		A split graph $G$ is called {\em $\pi$-convex with convexity on $K$} if there is an associated structure $\pi$ on $K$ such that for each $v\in I$, $N_G(v)$ induces a connected subgraph in $\pi$.
	\end{definition}
	\begin{definition}
		A split graph $G$ is called {\em $\pi$-convex with convexity on $I$} if there is an associated structure $\pi$ on $I$ such that for each $v\in K$, $N^I_G(v)$ induces a connected subgraph in $\pi$.
	\end{definition}
	\noindent In general $\pi$ can be any arbitrary structure. In this paper, We consider the following structures for $\pi$; "tree", "star", "comb", "path", "triad", and "cycle".  Note that the structure $\pi$ in $G$ is an imaginary structure.\\
	In the rest of the sections, we solve STREE for the case $R=I$ and it is sufficient to look at this case and all other cases can be solved using $R=I$ as a black box.  In Section \ref{ocases}, we present a transformation using which we can solve other cases.
	\section{The classical complexity of STREE} \label{ccstree}
	In Section \ref{sci}, we analyze the classical complexity of STREE on split graphs with convexity on $I$, and in Section \ref{sck}, we analyze the classical complexity of STREE on split graphs with convexity on $K$.
	\subsection{STREE in split graphs with convexity on $I$}\label{sci}
	%We consider tree-convex, star-convex, comb-convex, path-convex, triad-convex and circular-convex split graphs. 
	%	In this section, we consider convex split graphs with convexity on $I$ and
	When we refer to convex split graphs in this section, we refer to convex split graphs with convexity on $I$.  For STREE on split graphs with convexity on I, we establish hardness results for star-convex and comb-convex split graphs, and polynomial-time algorithms for path-convex, triad-convex, and circular-convex split graphs.
	\subsubsection{Star-convex split graphs}\label{ssi}
	In this section, we establish a classical hardness of STREE on star-convex split graphs by presenting a polynomial-time reduction from the Exact-3-Cover problem to STREE on star-convex split graphs.\\
	The decision version of Exact-3-Cover problem (X3C) is defined below:
	\begin{center}
		\fbox{\parbox[c][][c]{0.95\textwidth}{    
				\emph{X3C $(X,\mathcal{C})$}
				\\\textbf{Instance:} A finite set $X=\{x_1,\ldots,x_{3q}\}$ and a collection $\mathcal{C}=\{C_1,C_2,\ldots,C_m\}$ of 3-element subsets of $X$.
				\\\textbf{Question:} 
				Is there a subcollection $\mathcal{C'}$ $\subseteq$ $\mathcal{C}$ such that for every $x \in X$, $x$ belongs to exactly one member of $\mathcal{C'}$ (that is, $\mathcal{C'}$ partitions $X$) ?
		} }	
	\end{center}
	\noindent The decision version of Steiner tree problem (STREE) is defined below:
	\begin{center}
		\fbox{\parbox[c][][c]{0.95\textwidth}{    
				\emph{STREE $(G,R,k)$}
				\\\textbf{Instance:} A graph $G$, a terminal set $R\subseteq V(G)$, and a positive integer $k$.	
				\\\textbf{Question:} Is there a set $S\subseteq V(G)\setminus R$ such that $|S|\leq k$, and $G[S\cup R]$ is connected ?
		} }	
	\end{center}	
	\begin{theorem}\label{sti}
		For star-convex split graphs, STREE is NP-complete.
	\end{theorem}
	\begin{proof}
		\textbf{STREE is in NP:} Given a star-convex split graph $G$ and a certificate $S\subseteq V(G)$, we show that there exists a deterministic polynomial-time algorithm for verifying the validity of $S$.  Note that the standard Breadth First Search (BFS) algorithm can be used to check whether $G[S\cup R]$ is connected.  It is easy to check whether $|S|\leq k$.  The certificate verification can be done in $O(|V(G)|+|E(G)|)$.  Thus, we conclude that STREE is in NP.
		\\\textbf{STREE is NP-Hard:} It is known \cite{garey1979guide} that X3C is NP-complete.  X3C can be reduced in polynomial time to STREE on star-convex split graphs using the following reduction.  We map an instance $(X,\mathcal{C})$ of X3C to the corresponding instance $(G,R,k)$ of STREE as follows: $V(G)=V_1\cup V_2$, $V_1=\{c_i\mid 1\leq i \leq m\}$, $V_2=\{x_1,x_2,\ldots,x_{3q},x_{3q+1}\}$, $E(G)=\{\{c_i,x_j\}\mid x_j\in C_i, 1\leq j\leq 3q, 1\leq i \leq m\} \cup \{\{x_{3q+1},c_i\}\mid 1\leq i \leq m\} \cup \{\{c_i,c_j\}\mid 1\leq i\le j \leq m\}$.  Let $R=V_2$, $k=q$.  Note that $G$ is a split graph with $V_1$ being a clique and $V_2$ being an independent set.  Now we show that $G$ is a star-convex split graph by defining an imaginary star $T$ on $V_2$:
		\\Let $V(T)=V_2$ and $E(T)=\{\{x_{3q+1},x_i\}\mid 1\leq i \leq 3q\}$.  We see that $x_{3q+1}$ is the root of the star $T$.
		%In the reduced instance, we have an imaginary star with $3q+1$ as the root of the imaginary star, and the remaining vertices of $V_2$ are the pendant vertices of the imaginary star, and hence, $G$ is a star-convex split graph with convexity on $I$ with $K=V_1$ forms a clique in $G$, $I=V_2$ forms an independent set in $G$.
		%\\An illustration is included in the appendix Figure \ref{fig2}.
		%\noindent\textbf{An illustration for STREE in star-convex split graphs with convexity on $I$}
		\\An illustration for X3C with $X=\{x_1,x_2,x_3,x_4,x_5,x_6\}$ and $\mathcal{C}=\{C_1=\{x_1,x_2,x_3\},C_2=\{x_2,x_3,x_4\},C_3=\{x_1,x_2,x_5\},C_4=\{x_2,x_5,x_6\},C_5=\{x_1,x_5,x_6\}\}$, and the corresponding graph $G$ with $R=I$, $k=2$ is shown in Figure \ref{fig2}.  Note that the imaginary star on $I$ with the root $x_7$ is also shown in Figure \ref{fig2}.  For this instance the solution to X3C is $\mathcal{C'}=\{C_2,C_5\}$, and the corresponding solution for graph $G$ is $S=\{c_2,c_5\}$.
		\begin{figure}[H]
			\begin{center}
				\includegraphics[scale=0.7]{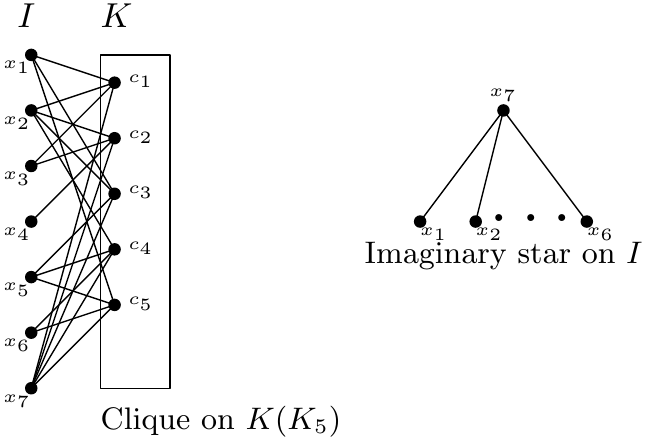}
				\caption{Reduction: An instance of X3C to STREE on star-convex split graphs}
				\label{fig2}
			\end{center}
		\end{figure}
		\begin{myclaim}
			$G$ is a star-convex split graph.
		\end{myclaim}
		\begin{proof}
			For each $c_i \in V_1$, $N_G^I(c_i)\subseteq V_2$.  By construction $x_{3q+1}$ is adjacent to all of $V_1$.  Therefore, for each $c_i\in K$, $N_G^I(c_i)$ is a subtree in $T$.  Hence $G$ is a star-convex split graph.
			\qed
		\end{proof}
		\begin{myclaim}\label{csti}
			Exact-3-Cover ($X,\mathcal{C}$) if and only if STREE ($G,R=V_2,k=q$)
			%($X,\mathcal{C}$) is a yes-instance of X3C if and only if ($G,R,\frac{|X|}{3}$) is a yes-instance of STREE.
		\end{myclaim}
		\begin{proof}
			\emph{Only if:} If there exists $\mathcal{C'}\subseteq\mathcal{C}$ which partitions $X$, then the set of vertices $S=\{c_i\in V_1\mid C_i\in \mathcal{C'}\}$, where $c_i$ is the vertex corresponding to $C_i$, forms a Steiner set with $R=V_2$.
			\\
			\emph{If:} Assume that there exists a Steiner tree $T$ in $G$ for $R=V_2$.  Let $S \subseteq V_1$ be the Steiner set of $T$, $|S|=q$.  We now construct the corresponding solution to X3C, $\mathcal{C'}=\{C_i\in C\mid c_i\in S\}$.  Since $|S|=q$, we have $|\mathcal{C'}|=q$.  Further, $S$ is the Steiner set for the terminal set $R=\{x_1,\ldots,x_{3q},x_{3q+1}\}$.  Therefore, for any $c_i\in S$, we have $|N^I_G(c_i)\setminus \{x_{3q+1}\}|=3$.  Since $|S|=q$ and $|I\setminus \{x_{3q+1}\}|=3q$, for all $c_i,c_j\in S, i\neq j$, $N^I_G(c_i)\cap N^I_G(c_j)=\{x_{3q+1}\}$.  Hence for all $ \{C_i,~C_j\}\subseteq C'$, we see that $C_i\cap C_j\neq \emptyset$.  Therefore, $\mathcal{C'}$ is the corresponding solution to X3C.
			%$S \subseteq V_1$ in $G$ on at most $k=q$ Steiner vertices, then observe that for all vertex $v \in S$, $d_{G}^I(v)=4$, $|S|= \frac{|V_2|-1}{3}$ and $|N_{G}^I(S)|=|V_2|$.  Note that $x_{3q+1}$ is adjacent to all the vertices in $V_2$, hence $|S|= \frac{|V_2|-1}{3}$.  It follows that there does not exist $u,v \in S$ such that $N_{G}^I(u) \cap N_{G}^I(v) \geq 2$.   Therefore, $\mathcal{C'}=\{c_i \in \mathcal{C} ~|~ v_i \in S\}$, where $v_i$ is the corresponding vertex of $c_i$, forms an exact-3-cover of $X$.
			\qed
		\end{proof}
		Thus we conclude STREE is NP-Hard on the star-convex split graph.  Therefore, STREE is NP-complete on star-convex split graphs.
		\qed
	\end{proof}
	\begin{corollary}
		For tree-convex split graphs, STREE is NP-complete.
	\end{corollary}
	\begin{proof}
		Since star-convex split graphs are a subclass of tree-convex split graphs, from Theorem \ref{sti}, this result follows. \qed
	\end{proof}
	%The study of parameterized complexity is concerned with designing algorithms with complexity $f(k)n^{O(1)}$, where $k$ is the parameter of interest, usually the solution size, and $n$ is the input size.% In \cite{dreyfus1971steiner} it is shown that STREE in general graphs is Fixed-parameter Tractable (FPT) if the parameter is the size of the terminal set.
	%It is known \cite{dom2009incompressibility} that STREE in general graphs with parameter $|S|$ is W[2]-hard. We now prove a similar result for our graph class.\\
	We next define the parameterized version of the Steiner tree problem and prove that Theorem \ref{sti} is indeed a parameter preserving reduction which we establish in Theorem \ref{pstree}.  Further, the following result strengthens the result of \cite{dom2009incompressibility}.\\
	\noindent The parameterized version of Steiner tree problem (PSTREE) is defined below:
	\begin{center}
		\fbox{\parbox[c][][c]{0.95\textwidth}{    
				\emph{PSTREE $(G,R,k)$}
				\\\textbf{Instance:} A star-convex split graph $G$, a terminal set $R\subseteq V(G)$.	
				\\\textbf{Parameter:} A positive integer $k$.
				\\\textbf{Question:} Is there a set $S\subseteq V(G)\setminus R$ such that $|S|\leq k$, and $G[S\cup R]$ is connected ?
		} }	
	\end{center}
	\begin{theorem}\label{pstree}
		For star-convex split graphs, STREE is W[1]-hard with parameter $|S|$.
	\end{theorem}
	\begin{proof}
		It is known \cite{ashok2015unique} that the parameterized Exact Cover problem (generalization of X3C) with parameter $|\mathcal{C'}|$ is W[1]-hard.  Note that the reduction presented in Theorem \ref{sti} maps $(X,\mathcal{C},q)$ to $(G,R,k=q)$.  From Claim 1.2 of Theorem \ref{sti}, we can observe that the reduction is a solution preserving reduction.  Hence the reduction is a deterministic polynomial-time parameterized reduction.  Therefore, PSTREE on star-convex split graphs is W[1]-hard.
		\qed
	\end{proof}
	Since the Steiner tree problem for $R=I$ on star-convex split graphs is unlikely to have a polynomial-time algorithm, we shall explore the following two subclasses of star-convex split graphs: (i) star-convex split graphs with bounded degree $d$ such that for each $y\in I,~d_G(y)\leq d$, and  (ii) star-convex split graph with imaginary star $T$ on $I$ with $l$ pendent vertices.  For (i), we present a polynomial-time algorithm, and for (ii), we present an FPT algorithm.  Let $T$ be the imaginary star on $I$.  In a graph $G$, the vertices $a,b\in V(G)$ are called twins, if $N_G[a]=N_G[b]$.  Observe that twins in a split graph can occur only in $K$.  For (i) and (ii), we consider graphs that do not have twins.  
	\\We shall now present a polynomial-time algorithm for star-convex split graphs with bounded degree $d$ such that for each $y\in I,~d_G(y)\leq d$.
	\begin{theorem}
		Let $G$ be a star-convex split graph with bounded degree $d$ such that for each $y\in I,~d_G(y)\leq d$.  A minimum Steiner tree $S$ can be found in polynomial time on $G$ for $R=I$.
	\end{theorem}
	\begin{proof}
		Let the root of $T$ be $z$.  By the structure of star-convex split graphs, we know that any $v\in K$ is either adjacent to $z$ or it is adjacent to exactly one vertex in $T$.  We consider the following two cases to find a minimum Steiner set of $G$ for $R=I$.
		\\\emph{Case 1:}  There exists $y$ in $(T-\{z\})$ such that $N_G(y)\cap N_G(z)= \emptyset$.
		\\Let $R_1=\{r\mid r\in (I\setminus \{z\}) \mbox{ such that } N_G(r)\cap N_G(z)= \emptyset\}$.  For each $r\in R_1$, we include the neighbor of $r$ in $S_1$, say $v$.  The set $S_1$ can be found in linear time.
		\\\emph{Case 2:}  There exists $y$ in $(T-\{z\})$ such that $N_G(y)\cap N_G(z)\neq \emptyset$.
		\\Let $R_2=\{s\mid s\in (I\setminus \{z\}) \mbox{ such that } N_G(s)\cap N_G(z)\neq \emptyset\}$.  Since $|N_G(z)|\leq d$, we find a minimum sized subset $S_2$ in $N_G(z)$ such that for each $s\in R_2$,  $N_G(s)\cap R_2\neq \emptyset$.  Since $d$ is a constant, the set $S_2$ can be found in linear time.
		\\If $R_1\neq \emptyset,~R_2\neq \emptyset$, then the $S$ of $G$ for $R=I$ is $S_1\cup S_2$.  If $S_2=\emptyset$, then the Steiner set $S$ of $G$ for $R=I$ is $S=S_1\cup \{v\}$, where $v \in N_G(z)$.  If $S_1=\emptyset$, then the Steiner set $S$ of $G$ for $R=I$ is $S=S_2$.  Observe that for each $a\in I$,  $N_G(a)\cap S\neq \emptyset$.   It is clear that $S$ is a Steiner set of $G$ for $R=I$.  
		\\
		For each vertex $r\in R_1$, $|N_G(r)\cap S|=1$, and for each vertex $s\in R_2$,  $|N_G(s)\cap S|=1$.  Note that $R_1\cap R_2=\emptyset$ and $R_1\cup R_2=I$.  Therefore, $S$ is a minimum Steiner set of $G$ for $R=I$.
		\qed
	\end{proof}
	Further, we analyze the complexity of STREE for $R=I$ on star-convex split graphs with the number of pendent vertices in the imaginary star is bounded, say $l$ (degree of root vertex in imaginary star $T$).  
	\noindent The parameterized version of the Steiner tree problem (PSTREE1) is defined below:
	\begin{center}
		\fbox{\parbox[c][][c]{0.95\textwidth}{    
				\emph{PSTREE1 $(G,R,k)$}
				\\\textbf{Instance:} A star-convex split graph $G$ with imaginary star $T$ on $I$ with $l$ pendent vertices, a terminal set $R=I$.	
				\\\textbf{Parameter:} A positive integers $l$ and $k$.
				\\\textbf{Question:} Is there a set $S\subseteq V(G)\setminus R$ such that $|S|\leq k$, and $G[S\cup R]$ is connected ?
		} }	
	\end{center}
	\begin{theorem}
		Let $G$ be an instance of PSTREE1.  Then $G$ has a kernel of size $2^l-1$.
	\end{theorem}
	\begin{proof}
		Let $z$ be the root of the imaginary star $T$.  Since $|V(T)|=l$, it is clear that $S\subseteq (N_G(V(T)\setminus \{z\}))$.  We preprocess the graph $G$ and $G'$ is obtained as follows; Let $Y=\{y\mid y\in I,~|N_G(y)|=1\}$.  We obtain the graph $G'=G-N_G[Y]$ with $k=k-|Y|$.  Let the imaginary structure in $I$ of $G'$ be $T'$.  The cardinality of $(N_{G'}(V(T')\setminus \{z\}))$ in $G'$ is at most $2^l-1$.  Thus $S'\subseteq N_{G'}(V(T)\setminus \{z\})$, and we obtain a kernel of size $2^l-1$ for PSTREE1.  From the kernel of size $2^l-1$, we obtain the Steiner set $S'$ of $G'$ by finding all possible subsets of size at most $k$.  Thus the Steiner set $S$ of $G$ for $R=I$ is obtained in time $O(2^{lk}n^2)$.
		\qed
	\end{proof}
	%It is known from Theorem \ref{sti}, that when the tree is a star, then STREE is NP-complete.  Further, it is interesting to look at some subclass $T'$ of the tree and analyze the complexity of split graph having $T'$ as its imaginary structure.  We shall look at the variant of tree which is comb and we shall analyze the classical complexity of STREE on comb-convex split graphs.
	\noindent\textbf{Highlights:}
	\\It turns out that STREE on tree-convex split graphs is NP-complete.  It is natural to ask for complexity when the imaginary tree has a special structure.  For example, binary-tree, ternary tree, and so on.  Interestingly, a comb is a special case of binary trees; in Section \ref{scs}, for comb-convex split graphs, we show that STREE is NP-complete.  As far as a study on unary-tree-convex split graphs is concerned, we observe that unary-tree-convex split graphs are precisely path-convex split graphs.  In Section \ref{pcs}, we show that STREE on path-convex split graphs is polynomial-time solvable.  This draws a thin line between P-versus-NPC input instances of STREE; polynomial-time solvable for unary-tree-convex split graphs and NP-complete for binary-tree-convex split graphs.
	One can also see the dichotomy status of this problem via these two structures as well.
	%We shall next consider comb-convexity and prove that STREE is NP-complete on comb-convex split graphs.
	\subsubsection{Comb-convex split graphs}\label{scs}
	We present a polynomial-time reduction from the vertex cover problem on general graphs to STREE on comb-convex split graphs.
	\\The decision version of Vertex Cover problem (VC) is defined below:
	\begin{center}
		\fbox{\parbox[c][][c]{0.95\textwidth}{    
				\emph{VC $(G,k)$}
				\\\textbf{Instance:} A graph $G$, a non-negative integer $k$.	
				\\\textbf{Question:} Does there exist a set $S\subseteq V(G)$ such that for each edge $e=\{u,v\}\in E(G)$, $u\in S$ or $v\in S$ and $|S|\leq k$ ?
		} }	
	\end{center}
	\begin{theorem}\label{ci}
		For comb-convex split graphs, STREE is NP-complete.
	\end{theorem}
	\begin{proof}
		%\textbf{STREE is in NP:}  Given an input instance $(G,R,k)$ of STREE and a certificate set $S\subseteq V(G)$, whether $S$ is a Steiner set of cardinality at most $k$ can be verified in polynomial time, as the connectedness of $G[R\cup S]$ can be verified in polynomial time by using standard graph traversal algorithms \cite{cormen2009introduction}.\\
		\textbf{STREE is NP-Hard:}  It is known \cite{cormen2009introduction} that VC on general graphs is NP-complete and this can be reduced in polynomial time to STREE on comb-convex split graphs using the following reduction.  We map an instance $(G,k)$ of VC on general graphs to the corresponding instance $(G^*,R,k'=k)$ of STREE as follows: $V(G^*)=V_1\cup V_2\cup V_3$, \\
		$V_1=\{x_i\mid v_i\in V(G)\}$, \\ $V_2=\{y_{i}\mid e_i\in E(G)\}$, \\
		$V_3=\{z_{i}\mid e_i\in E(G)\}$.\\
		We shall now describe the edges of $G^*$, \\
		$E(G^*)=E_1 \cup E_2\cup E_3$.  Let $n=|V(G)|,~m=|E(G)|$ \\
		$E_1= \{\{y_{i},x_p\},\{y_{i},x_q\},\mid e_i=\{v_p,v_q\}\in
		E(G),~x_p,x_q\in V_1,~y_{i}\in V_2,~1\leq i \leq m,~1\leq p \leq
		n,~1\leq q \leq n\} \\
		E_2=\{\{x,z_{i}\}\}\mid x\in V_1,~z_{i}\in V_3,~1\leq i \leq m\}$\\
		$E_3=\{\{x_i,x_j\}\mid, x_i,x_j \in V_1, 1\leq i \le j \leq n\}$.
		\\We define $K=V_1$, $I=V_2\cup V_3$, and the imaginary comb $T$ on $I$ is defined with $V_3$ as the backbone and $V_2$ as the pendant vertex set.  That is, $V(T)=I$ and $E(T)=\{\{y_{1},z_{1}\}, \{y_{2},z_{2}\}, \ldots, \{y_{i},z_{i}\}  \mid 1 \leq i \leq m \}$.
		%\\An illustration is included in Appendix Figure \ref{comb}.
		\\An example is illustrated in Figure \ref{comb},  the vertex cover instance $G$ with $k=2$ is mapped to STREE instance of comb-convex split graph $G^*$ with $R=\{y_{1},y_{2},y_{3}\}$, $k'=2$. 
		\begin{figure}[H]
			\begin{center}
				\includegraphics[scale=0.7]{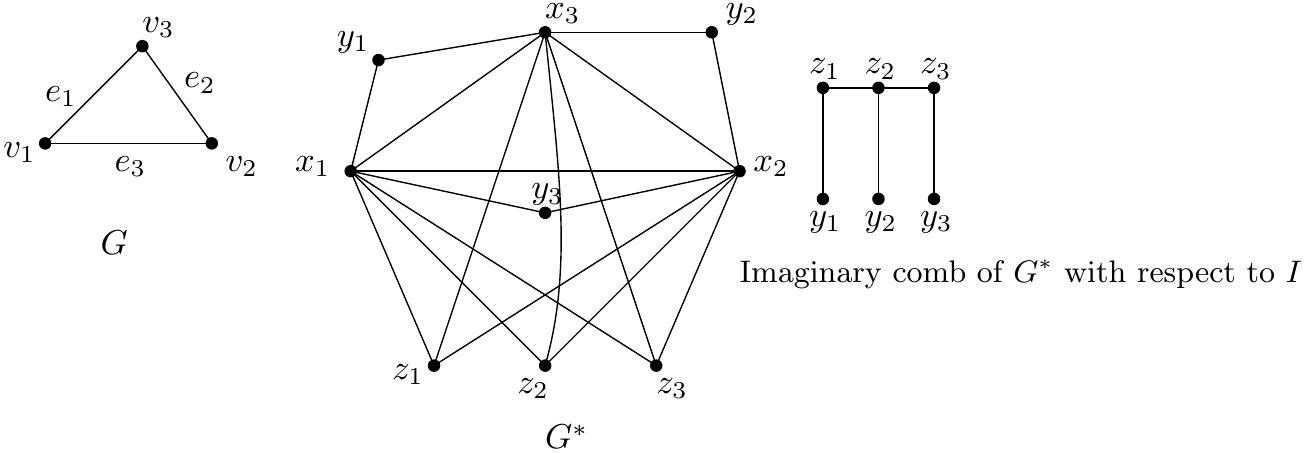}
				\caption{An example: VC reduces to STREE.}\label{comb}
			\end{center}
		\end{figure}
		\begin{myclaim}
			$G^*$ is a comb-convex split graph.
		\end{myclaim}
		\begin{proof}
			For each $x_i \in V_1$, $N_G^I(x_i)=V_3\cup W,~W\subseteq V_2$.  By our construction $x_i$ is adjacent to all of $V_3$.  Therefore, the graph induced on $N_G^I(x_i)$ is a subtree in $T$.  Hence $G^*$ is a comb-convex split graph.
			\qed
		\end{proof}
		\begin{myclaim}
			$(G,k)$ has a vertex cover with at most $k$ vertices if and only if
			$(G^*,R=\{y_{i}\mid 1\leq i\leq m\}\},k'=k)$ has a Steiner tree of size at most $k'=k$ Steiner vertices.
		\end{myclaim}
		\begin{proof}
			\emph{(Only if)} Let $V'=\{v_i\mid 1\leq i \leq k\}$ is a vertex cover of size $k$ in $G$.  Then we construct the Steiner set $S$ of $G^*$ for $R=\{y_{i}\mid 1\leq i\leq m\}$ as follows: $S=\{x_i \mid 1 \leq i \leq k, ~v_i \in V', ~x_i \in V(G^*)\}$.
			Since $V'$ is a vertex cover, for any edge $e_i=\{v_p,v_q\}\in E(G)$, $v_p$ or $v_q$ is in $V'$.  Hence $S$ contains $x_p$ or $x_q$.  Therefore, by the definition of $V_2$, for each vertex $y_i$, there exists a neighbor in $S$.  Since $V_1$ is a clique by our construction, $G[R\cup S]$ is connected. 
			%Indeed, for any edge $e_i=\{v_k,v_l\}\in E(G)$, $v_k$ or $v_l$ in $V'$.   
			%Then by our construction, we know that $y_{i}\in G^*$ corresponding to $e_i\in G$ is adjacent to $v_k$ and $v_l$, and  $x_k$ or $x_l$ is in $S$.  So each vertex in $\{y_{i}~|~1\leq i \leq m\}$ is adjacent to at least one vertex in $S$.   Further,  by our construction, each vertex in $V_1$ is adjacent to each vertex in $V_3$.  Hence $S \cup R$ induces a connected subgraph in $G^*$.
			\\\emph{(If)} For $R$ in $G^*$, let $S=\{x_i\mid 1\leq i \leq k'\}$ is a Steiner set of $G^*$ of size $k'$.  Then, we construct the vertex cover $V'$ of size $k$ in $G$ as follows; $V'=\{v_i\mid x_i \in S,~v_i\in V(G),~1\leq i\leq k'\}$.   We now claim that $V'$ is a vertex cover in $G$.   Suppose that there is an edge $e_i=\{v_p,v_q\}\in E(G)$ for which neither $v_p$ nor $v_q$ is in $V'$.   This implies that neither $x_p$ nor $x_q$ is in $S$.   Since $R$ contains $y_{i}$, it follows that $N_G(y_{i}) \cap S=\emptyset$.   Thus $S$ is not a Steiner set.  A contradiction.   Therefore, $V'$ is a vertex cover of size $k$ in $G$.
			\qed
		\end{proof}
		Thus we conclude STREE is NP-Hard.  Therefore, STREE is NP-complete on comb-convex split graphs.
		\qed	
	\end{proof}
	Having arbitrary comb $T$ as imaginary structure, STREE on comb-convex split graphs is NP-complete.  
		%A natural problem that can arise is "what happens when the imaginary structure is having a backbone path of length $l$ ?".  
		We show that finding STREE on comb-convex split graphs $G$ with backbone path of length $l$ for $R=I$ is in XP with respect to the parameter $l$.  Let the backbone path $B$ of imaginary comb $T$ be $(a_1,a_2\ldots,a_l)$.  Observe that $d_T(a_i)=3,~2\leq i \leq l-1$.  Hence Lemma \ref{ti} in triad-convex split graphs is also true for each $a_i\in I,~2\leq i \leq l-1$ in the comb-convex split graphs.
		\begin{lemma}
			Let $G$ be a comb-convex split graph.  Let $S$ be a minimum Steiner set of $G$ for $R=I$.  Then $1\leq |N_G(z)\cap S|\leq 3$, where $z\in \{a_2,\ldots,a_{l-1}\}$.
		\end{lemma}
		\begin{proof}
			The proof is similar to the proof of Lemma \ref{ti}.
		\end{proof}
		\noindent The parameterized version of the Steiner tree problem (PSTREE2) is defined below:
		\begin{center}
			\fbox{\parbox[c][][c]{0.95\textwidth}{    
					\emph{PSTREE2 $(G,R,k)$}
					\\\textbf{Instance:} A star-convex split graph $G$ with imaginary comb $T$ on $I$ with $l$ vertices in the backbone path, a terminal set $R=I$.	
					\\\textbf{Parameter:} A positive integer $l$.
					\\\textbf{Question:} Is there a set $S\subseteq V(G)\setminus R$ such that $|S|\leq k$, and $G[S\cup R]$ is connected ?
			} }	
		\end{center}
		\begin{theorem}
			Let $G$ be an instance of PSTREE2.  Then $S$ can be found in $O(n^{3l})$ time.
		\end{theorem}
		\begin{proof}
			Our proof is constructive.  For each $a_i\in I,~2\leq i \leq l-1$, we explore at most three vertex combinations in $N_G(a_i)$, which is similar to Algorithm \ref{algo2}.  The proof of correctness is similar to the proof of Theorem \ref{tcsi}.  Since there are $l$ vertices in the backbone path of $T$, $S$ can be found in $O(n^{3l})$ time.
			\qed
		\end{proof}
	%\begin{corollary}
	%	STREE is NP-complete on caterpillar-convex split graphs with convexity on $I$.
	%\end{corollary}
	%\begin{proof}
	%	Since comb-convex split graphs are subclass of caterpillar-convex split graphs, from Theorem \ref{ci} this result follows.
	%\end{proof}
	\noindent
	\textbf{Insights into reduction instances of Theorem \ref{ci}}
	\\	
	A closer look at the reduction instances of Theorem \ref{ci} reveals that the presence of pendant vertices in the comb makes the problem NP-hard.  It is natural to ask for the complexity of STREE in a variant of comb-convex split graphs where there are no pendant vertices (no teeth) in the comb which is precisely the class of path-convex split graphs.  Interestingly, STREE on path-convex split graphs is polynomial-time solvable, which we prove in the next section.
	%Since STREE on a comb-convex split graph with convexity on $I$ is NP-complete, we look in to a graph class where there is no pendant vertices in the imaginary comb.  Then it is a path-convex split graph with convexity on $I$.  We shall look into the complexity of STREE in path-convex split graph with convexity on $I$ in next section.
	\subsubsection{Path-convex split graphs}\label{pcs}
	In this section, we propose a polynomial-time algorithm for STREE on path-convex split graphs.  Recall that a split graph $G$ is called path-convex if there exists a linear ordering $\sigma$ of vertices in $I$ such that for each $u\in K$, $N_G(u)$ is consecutive in $I$ with respect to $\sigma$.
	\\
	Let $G$ be a path-convex split graph.  Let the vertices in $K$ be $w_1,\ldots,w_m$, and the vertices in $I$ be $x_1,\ldots,x_n$.  Path-convex split graphs can also be interpreted as follows: there exists an imaginary path $P=(x_1,\ldots,x_n)$ on $I$ such that for each $u\in K$, $N^I_G(u)$ is an interval (subpath in the imaginary path) in $I$.  When we refer to $x_i$ in $\sigma$, the index of $x_i$ in $\sigma$ is $i$.
	For $u\in K$, if $N_G^I(u)=\{x_p,\ldots,x_q\}$, then $l(u)=x_p$ and $r(u)=x_q$.  That is $l(u)$ is the least indexed vertex of $N^I_G(u)$ in $\sigma$, and $r(u)$ is the greatest indexed vertex in $N^I_G(u)$.
	%For each $u\in K$, $l(u)$ is the least indexed vertex of $N^I_G(u)$ in $\sigma$, and $r(u)$ is the greatest indexed vertex in $N^I_G(u)$.  If $N_G^I(u)=\{x_p,\ldots,x_q\},~u\in K$, then $l(u)=x_p$ and $r(u)=x_q$.
	For each $x_i\in I$, $1\leq i \leq n$, let $\alpha(x_i)=u$ such that $u\in N_G(x_i)$ and $r(u)$ is maximum.
	%Let $\alpha(x_i)$ represent $u\in K$ for which $r(u)$ is having maximum index and $u\in N_G(x_i)$. 
	For $w_i,w_j\in K$, when we write $w_i\prec w_j$, we mean that $r(w_i)$ appears before $r(w_j)$ with respect to $\sigma$.  We order the vertices in $K$ as follows; $w_1\prec w_2\prec \ldots \prec w_m$. 
	%We order the vertices in $K$ according to their right endpoints.  By $r(w_i)\prec r(w_j),~w_i,w_j\in K$, we mean that $r(w_i)$ appears before $r(w_j)$.
	%,and $r(w_j)\succ r(w_i)$, $w_i,w_j\in K$, we mean that $r(w_j)$ appears after $r(w_i)$.  
	\\
	The idea behind our Algorithm \ref{apci} is to visualize the neighborhood of each vertex in $K$ as intervals and each vertex in $I$ as points.  All points in $I$ are unmarked initially.  Choose the largest interval, say $\gamma$ starting from $x_1$.  Mark all the points in $I$ that are in $\gamma$.  Among the unmarked points in $I$ choose the point whose index is minimum, say $x_j$.  We continue our algorithm by choosing the interval, say $\beta$ that contains $x_j$ and whose right endpoint is maximum.  Mark all points in $I$ that are contained in $\beta$ and proceed in the similar line until we hit the point $x_t$.  This greedy approach is indeed optimum, which we establish in this section.
	%The idea behind our Steiner tree algorithm for path-convex split graphs is that it identifies $u\in K$ adjacent to $x_1$ such that $r(u)$ is maximum and we continue this from $r(u)+1^{th}$ vertex. This greedy approach is indeed optimum, which we establish in this section.
	%\\The following algorithm computes a minimum Steiner set for $G$.
	\begin{algorithm}[H]
		\caption{{\em STREE for path-convex split graphs.}}
		\label{apci}
		\begin{algorithmic}[1]
			\State{{\tt Input:}  A connected path-convex split graph $G$ and $R=I$. }
			\State{All vertices in $I$ are unmarked initially.  Let $i=1$, $b=r(\alpha(x_1))$, the Steiner set $S=\{\alpha(x_1)\}$.}
			%\State{All vertices in $I$ are unmarked initially.}
			%\State{Let $a=x_1,~S=\{\}$, $x_j=x_1$.}
			%\State{Initialize Steiner set $S=\{\alpha(x_1)\}$.}
			\State{Mark all vertices in $I$ that are adjacent to $\alpha(x_1)$.}
			%\State{Initialize $j$ as the index of $r(\alpha(a_1))$ in $\sigma$.}
			\While{$b\neq x_t$}
			%\State{Update $i=i+1$, $a_i=x_{j+1}$.}
			\State{Let $c$ be the least indexed unmarked vertex in $I$.}
			%\State{Update $b=r(\alpha(a_{i}))$, and $S=S\cup \{\alpha(b)\}$.}
			%\State{Update $j$ as the index of $r(\alpha(a_{i}))$ in $\sigma$.}
			\State{$b=r(\alpha(c))$, $S=S\cup\{\alpha(c)\}$.}
			\EndWhile
		\end{algorithmic}
	\end{algorithm}
	\noindent Let $S=\{u_1,\ldots, u_p\}$ be the Steiner vertices chosen by the algorithm.  Note that as per our algorithm $u_1\prec u_2\prec \ldots \prec u_p$.  Let $S'=\{u'_1,\ldots, u'_q\}$ be the Steiner vertices chosen by any optimal algorithm.  Without loss of generality, we arrange $S'$ such that $u'_1\prec u'_2\prec \ldots \prec u'_q$.  Since $R=I$, observe that $S\subseteq K$ and $S'\subseteq K$.
	%Let the Steiner vertices chosen by the algorithm $S$ be $ $ such that $r(u_1)\preceq r(u_2)\preceq \ldots \preceq r(u_p)$, and let the Steiner vertices chosen by an optimal solution $S'$ be $v_1,\ldots,v_q$ such that $r(v_1)\preceq r(v_2)\preceq \ldots \preceq r(v_q)$.  By $r(w_i)\preceq r(w_j),~w_i,w_j\in K$, we mean that the vertex $r(w_i)$ appears before $r(w_j)$.
	\begin{theorem}\label{tpci}
		For all indices $i\leq q$, $N_G^I(\{u_1,\ldots,u_i\})\supseteq N_G^I(\{u'_1,\ldots,u'_i\})$
	\end{theorem}
	\begin{proof}
		By mathematical induction on $i$, $i\geq  1$.
		\\\textbf{Base Case:} For $i=1$, 
		\\Since $u'_1\prec u'_j,~j>1$, we have $\{x_1,u'_1\}\in E(G)$.  Since our algorithm has chosen $u_1$, $\{x_1,u_1\}\in E(G)$ and by Step 2 of Algorithm \ref{apci}, $u_1=\alpha(x_1)$.  Therefore, $u'_1\prec u_1$.  The ordering of $K$ and the convexity on $I$ imply that $N_G(u_1)\supseteq N_G(u'_1)$.
		\\\textbf{Induction Hypothesis:} Assume for $i\geq 2$, $N^I_G(\{u_1,\ldots,u_{i-1}\}) \supseteq N^I_G(\{u'_1,\ldots,u'_{i-1}\})$ is true.
		\\\textbf{Induction Step:} We prove that when $i\geq 2$, $N^I_G(\{u_1,\ldots,u_i\})\supseteq N^I_G(\{u'_1,\ldots,u'_i\})$.
		\\By the induction hypothesis, we know that up to $i-1$, $N^I_G(\{u_1,\ldots,u_{i-1}\}) \supseteq N^I_G(\{u'_1,\ldots,u'_{i-1}\})$.  Observe that as per Step 6 of the algorithm, we have included $u_i\in S$.  This implies that after inclusion of $u_1,\ldots u_{i-1}$ into the solution, $u_i$ refers to $\alpha(c)$ where $c$ is the least indexed unmarked vertex in $I$.
		\\Assume on the contrary, $N^I_G(\{u_1,\ldots,u_i\})\nsupseteq N^I_G(\{u'_1,\ldots,u'_i\})$.  Then there exists $y\in I$ such that $y\in N_G(u'_i)$ and $y\notin N_G(u_i)$.  It is clear that $\alpha(y)\prec u_i$.  Since our algorithm must have included at least one vertex adjacent to $y$, it must be the case that $y\in N_G(u_j)$, for some $j$, $1\leq j \leq i-1$.  This implies that $y\in N^I_G(\{u_1,\ldots,u_{i-1}\})$, which is a contradiction.
		\\Therefore, $N_G^I(\{u_1,\ldots,u_i\})\supseteq N_G^I(\{u'_1,\ldots,u'_i\})$.  Hence the proof.
		\qed
	\end{proof}
	\begin{theorem}\label{tpi}
		Algorithm \ref{apci} outputs a minimum Steiner set, that is $p=q$.
	\end{theorem}
	\begin{proof}
		By Theorem \ref{tpci}, we know that $N_G^I(\{u_1,\ldots,u_q\})\supseteq N_G^I(\{u'_1,\ldots,u'_q\})$, and hence $|S|\leq |S'|$. Since $S'$ is an optimal solution, $|S|\geq |S'|$.  Therefore, $|S|=|S'|$, and $p=q$.
		\qed
	\end{proof}
	It is easy to see that Algorithm \ref{apci} runs in time $O(mn)$.\\
	%The greedy algorithm for a path-convex split graph with convexity on $I$ runs in time $O(mn)$.\\
	%Note that the time complexity of path-convex split graph with convexity on $I$ algorithm is $O(nm)$, since 
	Now we see that for comb-convex split graphs, STREE is NP-complete, whereas, for path-convex split graphs, STREE is polynomial-time solvable.
	This brings out the P-versus-NPC investigation of STREE on tree-convex split graphs.  This is one of the objectives of this research.
	\\It is important to highlight that we can solve STREE on triad-convex and circular-convex split graphs by using the algorithm of STREE on path-convex split graphs as a black box, which we prove in the following two sections.  
	\\\\
	\textbf{Application 1: Triad-convex split graphs}
	\\\\
	We investigate the classical complexity of STREE on triad-convex split graphs which are a variant of path-convex split graphs.  Since a triad structure has three paths with a common endpoint (root vertex), we shall explore the possibility of solving STREE on triad-convex split graphs using the algorithm for STREE on path-convex split graphs as a black box.
	\\
	We now present a polynomial-time reduction to map the instances of triad-convex split graphs to the instances of path-convex split graphs.  The reduction is similar to the reduction presented in \cite{pandey2019domination}. 
	%To prove that STREE on triad-convex split graph is polynomial-time solvable, we give a polynomial-time reduction from triad-convex split graph to path-convex split graph.
	\\
	Let $G$ be a triad-convex split graph with triad $T$ defined on $I$ such that for every $u\in K, ~N_G^I(u)$ is a subtree in $T$.  Let $z$ be the root vertex of the triad $T$.  There are three paths in $T-z$, let those paths be $B_1,B_2,$ and $B_3$.  Let the vertices in $B_i$ be $(x^i_1,~x^i_2,\ldots,~x^i_{|B_i|}),~1\leq i \leq 3$.  Let the vertices in $K$ be $w_1,\ldots,w_m$.
	%Let the vertices in $B_1$ be $x_{11},x_{12},\ldots,x_{1a}$, the vertices in $B_2$ be $x_{21},x_{22},\ldots,x_{2b}$, and the vertices in $B_3$ be $x_{31},x_{32},\ldots,x_{3c}$, $a\geq 1,~b\geq 1,~c\geq 1$ such that 
	Since $T$ is a triad, $N_T(z)=\{x_1^1, x_1^2, x_1^3\}$.  Let $N_G(z)=\{u_1,\ldots,u_s\}$.  Observe that $N_G(z)\subseteq K$.
	\\For each vertex $u_j\in N_G(z)$, $1\leq j \leq s$, we define $r(B_i,u_j)=x^i_p$ such that $N_G(u_j)\cap B_i\neq \emptyset$, $x^i_p\in N_G(u_j)$ and there does not exist $x^i_q\in N_G(u_j)$, $p+1\leq q\leq |B_i|$.  Let 
	\begin{equation*}
		\alpha_{i,j}=\max_{u_j}~r(B_i,u_j),~\beta_{i,j}=v,~v\in (N_G(\alpha_{i,j})\cap N_G(z)), ~1\leq i \leq 3,~1\leq j\leq s.
	\end{equation*}
	The following lemma is a key result for our reduction in work.
	\begin{lemma}\label{ti}
		Let $G$ be a triad-convex split graph.  Let $S$ be a minimum Steiner set of $G$ for $R=I$.  Then, $1\leq |N_G(z)\cap S|\leq 3$, where $z$ is the root of the triad.
	\end{lemma}
	\begin{proof}
		It is clear that for every vertex $y\in R$, $|N_G(y)\cap S|\geq 1$.
		\\
		Suppose that $|N_G(z)\cap S|>3$.  We claim that $N^I_G(\beta_{1,j})\cup N^I_G(\beta_{2,k})\cup N^I_G(\beta_{3,l})\supseteq N^I_G(N_G(z)\cap S),~1\leq j< k < l \leq s$.  
		On the contrary, there exists $y\in I$ such that $y\in N_G^I(N_G(z)\cap S)$ and $y\notin (N^I_G(\beta_{1,j})\cup N^I_G(\beta_{2,k})\cup N^I_G(\beta_{3,l}))$.  Without loss of generality, we shall assume that $y\in B_1$.  Let $\alpha_{1,j}$ be $x^1_p$, for some $p$, $1\leq p \leq |B_1|$.  Then $y\in \{x^1_1,\ldots,x^1_p\}$.  By the definition of $\beta_{1,j}$, we know that $\{x^1_1,\ldots,x^1_p\}\subseteq N^I_G(\beta_{1,j})$.  Hence, $y\in (N^I_G(N_G(z)\cap S))\cap B_1$ and $y\in N_G^I(\beta_{1,j})$, which is a contradiction that $y\notin N_G^I(\beta_{1,j})$.  Similarly, the argument is true if $y\in \beta_{2,k}$ or $y\in\beta_{3,l}$.  Thus there exists $y\in I$ such that $y\in N_G^I(N_G(z)\cap S)$ and $y\notin (N^I_G(\beta_{1,j})\cup N^I_G(\beta_{2,k})\cup N^I_G(\beta_{3,l}))$ is a contradiction and $N^I_G(\beta_{1,j})\cup N^I_G(\beta_{2,k})\cup N^I_G(\beta_{3,l})\supseteq N^I_G(N_G(z)\cap S)$.
		\\
		Observe that $\beta_{1,j},~\beta_{2,k},~\beta_{3,l}$ need not be distinct always.  %Since $S$ is a minimum Steiner set of $G$ with $R=I$ and, $|N_G(z)\cap S|>3$, then 
		Consider $S'=(S\setminus N_G(z))\cup \{\beta_{1,j},~\beta_{2,k},~\beta_{3,l}\}$.  Note that $|S'|\leq |S|-1$, which is a contradiction that $S$ is a minimum Steiner set of $G$ for $R=I$.
		\qed
	\end{proof}
	The above lemma indicates that $|N_G(z)\cap S|=1$, $|N_G(z)\cap S|=2$, and $|N_G(z)\cap S|=3$.  Accordingly, for each triad-convex split graph, we construct a corresponding set of path-convex split graphs as part of Construction 1, Construction 2, and Construction 3 which are explained below.  Further, using our construction and Algorithm \ref{apci}, we obtain a polynomial-time algorithm for triad-convex split graphs.  We have the following cases:
	\\\\
	Case 1: Exactly one neighbor of $z$ is in $S$.
	\\
	Case 2: Exactly two neighbors of $z$ is in $S$.
	\\
	Case 3: Exactly three neighbors of $z$ is in $S$.
	\\\\
	For each $u_j\in N_G(z),~1\leq j\leq s$, $A(u_j)$ represents a minimum Steiner set of $G$ for $R=I$ containing $u_j$.  For each $u_j,u_k\in N_G(z),~1\leq j< k\leq s$, $A(u_j,u_k)$ represents a minimum Steiner set of $G$ for $R=I$ containing $u_j,u_k$.  For each $u_j,u_k,u_l,~\{u_j,u_k,u_l\}\subseteq N_G(z),~1\leq j <k<l\leq s$, $A(u_j,u_k,u_l)$ represents a minimum Steiner set of $G$ for $R=I$ containing $u_j,u_k,u_l$.
	%Let $A(u_j),~1\leq j \leq s$ denotes a minimum Steiner set of $G$ with $R=I$ and containing exactly one neighbor of $z$, $u_j$.  We define $S_1=\{A(u_j)\mid u_j\in N(z),~1\leq j \leq s\}$.  Let $A(u_j,u_k),~1\leq j<k \leq s$ denotes the minimum Steiner set of $G$ when $R=I$ and containing exactly two neighbors of $z$, say $u_j,~u_k,~1\leq j<k \leq s$.  We define $S_2=\{A(u_j,u_k)\mid u_j,u_k\in N(z),~1\leq j<k \leq s\}$.  Let $A(u_j,u_k,u_l),~1\leq j<k<l \leq s$ denotes the minimum Steiner set of $G$ when $R=I$ and containing exactly three neighbors of $z$, say $u_j,~u_k,~u_l,~1\leq j<k<l \leq s$.  We define $S_3=\{A(u_j,u_k,u_l)\mid \{u_j,u_k,u_l\}\subseteq N(z),~1\leq j<k<l\leq s\}$.
	\\
	\textbf{Computation of a minimum Steiner set in each of the three cases.}
	\\
	For each case, we shall construct a set of path-convex split graphs using which we compute a minimum Steiner set $S$ of $G$.
	\\
	\textbf{Case 1: Exactly one neighbor of $z$ is in $S$.}
	\\\textbf{Construction 1:}
	\\For each $u_j\in N_G(z),~1\leq j \leq s$, we do the following:
	\\Since $(N_G(z)\setminus \{u_j\})\cap S=\emptyset$, using the graph $G_j=G-(N_G(z)\cup N^I_G(u_j))$, we construct $G^i_j$, for each $i,~1\leq i\leq 3$ as follows;
	\\ The graph $G^i_{j}$ with $I^i_j=B_i\cap(I\setminus\{N^I_G(u_j)\})$,~$K^i_j=K\cap (N_G(B_i)\setminus N_G(z))$, and $E(G^i_{j})=\{\{x,y\}\mid x,y\in V(G^i_{j}),~\{x,y\}\in E(G)\}$.
	\\The ordering of $I^i_j$ for $G^i_j$ is given by $B_i\cap(I\setminus\{N^I_G(u_j)\})$ with respect to the ordering $I$ in $G$.  Hence each $G^i_j$ is a path-convex split graph with ordering on $I^i_j$.
	Let $S^i_j$ be a minimum Steiner set $G^i_{j}$ for $R=I^i_j$, obtained using Algorithm \ref{apci}.  Then a minimum Steiner set of $G$ for $R=I$ containing $u_j$ is $A(u_j)=S^i_j\cup \{u_j\}$ and the proof for minimality of $A(u_j)$ is established in Lemma \ref{lti1}.
	\begin{lemma}\label{lti1}
		For some $u_j\in N_G(z)$, if $N_G(z)\cap S=\{u_j\}$ then $|S|=|S^1_j|+|S^2_j|+|S^3_j|+1$.
		%For $1\leq j \leq s$, if $S^i_j,~1\leq i \leq 3$ are minimum Steiner sets of $G^i_{j},~1\leq i \leq 3$ when $R=I$, then $\bigcup\limits_{i=1}^3 S^i_j\cup \{u_j\}$ is a Steiner set of $G$ when $R=I$.  Moreover, if there exists a minimum Steiner set, say $S$ of $G$ when $R=I$ such that $N_G(z)\cap S=\{u_j\}$, then $|S|=|S^1_j|+|S^2_j|+|S^3_j|+1$.
	\end{lemma}
	\begin{proof}
		Let $S^1_j\cup S^2_j\cup S^3_j,~1\leq j \leq s$ be a Steiner set of the graph $G^1_{j}\cup G^2_{j}\cup G^3_{j}$.  Let $G'$ be the graph induced on $N^I_G[u_j] \cup V(G^1_{j})\cup V(G^2_{j})\cup V(G^3_{j}),~1\leq j \leq s$.  The graph $G'$ for $R=I$, the Steiner set is $S'=\{u_j\}\cup S^1_j\cup S^2_j\cup S^3_j$.  We can observe that $G'$ is a subgraph of $G$, since $V(G)=V(G')\cup N_G(z)$.  Thus $S'$ is a Steiner set of $G$ for $R=I$.  For any minimum Steiner set $S$ of $G$ containing $u_j$ for $R=I$, clearly, $|S|\leq |S'|= |S^1_j|+|S^2_j|+|S^3_j|+1$.
		\\\\
		We now show that for any minimum Steiner set $S$ of $G$ containing $u_j$ for $R=I$, $|S|\geq |S^1_j|+|S^2_j|+|S^3_j|+1$.
		Assume $S$ is the minimum Steiner set of $G$ for $R=I$ and $N_G(z)\cap S=\{u_j\}$.  Then $S^1_j$ is a Steiner set of $G^1_{j}$ for $R=I^1_j$ such that $S^1_j=S\cap V(G^1_{j})$.  For each $v\in V(G^1_{j})$, $N_{G^1_{j}}(v)\cap S^1_j\neq \emptyset$.  Similarly, for $R=I^2_j$ of $G^2_{j}$, $S^2_j=S\cap V(G^2_{j})$ and for $R=I^3_j$ of $G^3_{j}$, $S^3_j=S\cap V(G^3_{j})$.  Hence,
		\\$S=(S\cap V(G^1_{j}))\cup (S\cap V(G^2_{j})) \cup (S\cap V(G^3_{j}))\cup \{u_j\}$
		\\$|S|=|(S\cap V(G^1_{j}))\cup (S\cap V(G^2_{j})) \cup (S\cap V(G^3_{j}))\cup \{u_j\}|$
		\\$=|(S\cap V(G^1_{j}))\cup (S\cap V(G^2_{j})) \cup (S\cap V(G^3_{j}))|+1$
		\\Since $|S\cap V(G^1_{j})|\geq |S^1_j|$, $|S\cap V(G^2_{j})|\geq |S^2_j|$,and $|S\cap V(G^3_{j})|\geq |S^3_j|$, we get
		\\$|S|\geq |S^1_j|+|S^2_j|+|S^3_j|+1$
		\\Therefore, $|S|=|S^1_j|+|S^2_j|+|S^3_j|+1$ is a minimum Steiner set of $G$ for $R=I$.
		\qed
	\end{proof}
	\noindent\textbf{Case 2: Exactly two neighbors of $z$ is in $S$.}
	\\\textbf{Construction 2:}
	\\For each $u_j,u_k\in N_G(z),~1\leq j<k \leq s$, we do the following:
	\\Since $(N_G(z)\setminus \{u_j,u_k\})\cap S=\emptyset$, using the graph $G_{jk}=G-(N_G(z)\cup N^I_G(u_j)\cup N^I_G(u_k))$, we construct $G^i_{jk}$, for each $i,~1\leq i \leq 3$ as follows;
	\\The graph $G^i_{jk}$ with $I^i_{jk}=B_i\cap(I\setminus\{N^I_G(u_j)\cup N^I_G(u_k)\})$,~$K^i_{jk}=K\cap (N_G(B_i)\setminus N_G(z))$, and $E(G^i_{jk})=\{\{x,y\}\mid x,y\in V(G^1_{ij}),~ \{x,y\}\in E(G)\}$.
	\\
	The ordering of $I^i_{jk}$ for $G^i_{jk}$ is given by $B_i\cap (I\setminus(\{N^I_G(u_j)\}\cup N^I_G(u_k)))$ with respect to ordering $I$ in $G$.  Hence each $G^i_{jk}$ is a path-convex split graph with ordering on $I^i_{jk}$.
	Let $S^i_{jk}$ a minimum Steiner set of $G^i_{jk}$ for $R=I^i_{jk}$ is obtained using Algorithm \ref{apci}.  Then a minimum Steiner set of $G$ for $R=I$ containing $u_j,~u_k$ is $A(u_j,u_k)=S^i_{jk}\cup \{u_j,u_k\}$ and the proof of minimality of $A(u_j,u_k)$ is as per Lemma \ref{lti2}.
	\begin{lemma}\label{lti2}
		For some $u_j,u_k\in N_G(z)$, if $N_G(z)\cap S=\{u_j,u_k\}$ then $|S|=|S^1_{jk}|+|S^2_{jk}|+|S^3_{jk}|+2$.
		%For $1\leq j<k \leq s$, if $S^1_{jk},~S^2_{jk},~S^3_{jk}$ are minimum Steiner sets of $G^1_{jk},~G^2_{jk},~G^3_{jk}$ when $R=I$, then $S^1_{jk}\cup S^2_{jk}\cup S^3_{jk}\cup \{u_j,u_k\}$ is a Steiner set of $G$ when $R=I$.  Moreover, if there exists a minimum Steiner set, say $S$ of $G$ when $R=I$ such that $N_G(z)\cap S=\{u_j,u_k\}$, then $|S|=|S^1_{jk}|+|S^2_{jk}|+|S^3_{jk}|+2$.
	\end{lemma}
	\begin{proof}
		The proof is similar to the proof of Lemma \ref{lti1}.
		\qed
	\end{proof}
	\noindent\textbf{Case 3: Exactly three neighbors of $z$ are in $S$.}
	\\\textbf{Construction 3:}
	\\For each $u_j,u_k,u_l$ such that $\{u_j,u_k,u_l\}\subseteq N_G(z),~1\leq j<k<l \leq s$, we do the following:
	\\Since $(N_G(z)\setminus \{u_j,u_k,u_l\})\cap S=\emptyset$, using the graph $G_{jkl}=G-(N_G(z)\cup N^I_G(u_j)\cup N^I_G(u_k)\cup N^I_G(u_l))$, we construct $G^i_{jkl}$, for each $i,~1\leq i \leq 3$ as follows;
	\\The graph $G^i_{jkl}$ with $I^i_{jkl}=B_i\cap(I\setminus\{N^I_G(u_j)\cup N^I_G(u_k)\cup N^I_G(u_l)\}$,~$K^i_{jkl}=K\cap (N_G(B_i)\setminus N_G(z))$, and $E(G^i_{jkl})=\{\{x,y\}\mid x,y\in V(G^i_{jkl}),~ \{x,y\}\in E(G)\}$.
	\\
	\\
	The ordering of $I^i_{jkl}$ for $G^i_{jkl}$ is given by $B_i\cap (I\setminus(\{N^I_G(u_j)\}\cup N^I_G(u_k)\cup N^I_G(u_l)))$ with respect to ordering $I$ in $G$.  Hence each $G^i_{jkl}$ is a path-convex split graph with ordering on $I^i_{jkl}$.
	Let $S^i_{jkl}$ a minimum Steiner set of $G^i_{jkl}$ for $R=I^i_{jkl}$ is obtained using Algorithm \ref{apci}.  Then a minimum Steiner set of $G$ for $R=I$ containing $u_j,~u_k,~u_l$ is $A(u_j,u_k,u_l)=S^i_{jkl}\cup \{u_j,u_k,u_l\}$ and the proof of minimality of $A(u_j,u_k,u_l)$ is as per Lemma \ref{lti3}.
	%\\Let $S$ be a minimum Steiner set of $G$ for $R=I$.  Based on the above constructions, we observe a few key lemmas which are presented next.
	\begin{lemma}\label{lti3}
		For some $u_j,u_k,u_l$ such that $\{u_j,u_k,u_l\}\subseteq N_G(z)$, If $N_G(z)\cap S=\{u_j,u_k,u_l\}$ then $|S|=|S^1_{jkl}|+|S^2_{jkl}|+|S^3_{jkl}|+3$.
		%For $1\leq j<k<l \leq s$, if $S^1_{jkl},~S^2_{jkl},~S^3_{jkl}$ are minimum Steiner sets of $G^1_{jkl},~G^2_{jkl},~G^3_{jkl}$ when $R=I$, then $S^1_{jkl}\cup~S^2_{jkl}\cup~S^3_{jkl}\cup \{u_j,u_k,u_l\}$ is a Steiner set of $G$ when $R=I$.  Moreover, if there exists a minimum Steiner set, say $S$ of $G$ when $R=I$ such that $N_G(z)\cap S=\{u_j,u_k,u_l\}$, then $|S|=|S^1_{jkl}|+|S^2_{jkl}|+|S^3_{jkl}|+3$.
	\end{lemma}
	\begin{proof}
		The proof is similar to the proof of Lemma \ref{lti1}.
		\qed
	\end{proof}
	\noindent Now we shall present an algorithm to find a minimum Steiner set of $G$ for $R=I$.
	%\\By using Algorithm \ref{algo2}, we compute the minimum Steiner set for triad-convex split graphs.
	\begin{algorithm}[H]
		\caption{{\em STREE for triad-convex split graphs.}}
		\label{algo2}
		\begin{algorithmic}[1]
			\State{{\tt Input:}  A connected triad-convex split graph $G$, $R=I$. }
			\State{Let $z$ be the central vertex of triad $T$, and let $S=\emptyset,~S_1=\emptyset,~S_2=\emptyset,~S_3=\emptyset$.}
			\ForAll{$u_j,~u_j\in N_G(z)$}
			\State{Construct $G^1_{j},~G^2_{j},~G^3_{j}$ using Construction 1.}
			\State{Using Algorithm \ref{apci}, find minimum Steiner sets $S^1_{j}$,  $S^2_j$, and $S^3_j$ for $G^1_{j},~G^2_{j},$ and $G^3_{j}$, respectively.}
			\State{Update $S_1=S_1\cup \{S^1_j\cup S^2_j\cup S^3_j\cup \{u_j\}\}$}
			\EndFor
			\ForAll{$u_j,u_k,~u_j,u_k\in N_G(z)$}
			\State{Construct $G^1_{jk},~G^2_{jk},~G^3_{jk}$ using Construction 2.}
			\State{Using Algorithm \ref{apci}, find minimum Steiner sets $S^1_{jk}$,  $S^2_{jk}$, and $S^3_{jk}$ for $G^1_{jk},~G^2_{jk},$ and $G^3_{jk}$, respectively.}
			\State{Update $S_2=S_2\cup \{S^1_{jk}\cup S^2_{jk}\cup S^3_{jk}\cup \{u_j,u_k\}\}$}
			\EndFor
			\ForAll{$u_j,u_k,u_l,~\{u_j,u_k,u_l\}\subseteq N_G(z)$}
			\State{Construct $G^1_{jkl},~G^2_{jkl},~G^3_{jkl}$ using Construction 3.}
			\State{Using Algorithm \ref{apci}, find minimum Steiner sets $S^1_{jkl}$,  $S^2_{jkl}$, and $S^3_{jkl}$ for $G^1_{jkl},~G^2_{jkl},$ and $G^3_{jkl}$, respectively.}
			\State{Update $S_3=S_3\cup \{S^1_{jkl}\cup S^2_{jkl}\cup S^3_{jkl}\cup \{u_j,u_k,u_l\}\}$}
			\EndFor
			\State{The minimum cardinality set in $S_1\cup S_2 \cup S_3$ is $S$.}
		\end{algorithmic}
	\end{algorithm}
	\noindent The proof of correctness of Algorithm \ref{algo2} follows from Lemmas \ref{ti}, \ref{lti1}, \ref{lti2}, and \ref{lti3}.  Observe that in Case 1 for $u_j\in N_G(z)$, the time required for constructing $G^1_j,~G^2_j$, and $G^3_j$ is $O(n^2)$.  Finding the Steiner set for each of $G^1_j,~G^2_j$, and $G^3_j$ incurs $O(n^2)$ time.  Thus finding the Steiner set for each $u_j\in N_G(z)$ incurs $O(n^3)$ time.  Similarly, for Case 2, the time required for constructing $G^1_{jk},~G^2_{jk}$, and $G^3_{jk}$ is $O(n^2)$.  Finding the Steiner set for each of $G^1_{jk},~G^2_{jk}$, and $G^3_{jk}$ incurs $O(n^2)$ time.  Thus the finding the Steiner set for each $u_j,u_k\in N_G(z)$ incurs $O(n^4)$ time.  Similarly, for the Case 3, the time required for constructing $G^1_{jkl},~G^2_{jkl}$, and $G^3_{jkl}$ is $O(n^2)$.  Finding the Steiner set for $G^1_{jkl},~G^2_{jkl}$, and $G^3_{jkl}$ incurs $O(n^2)$ time.  Thus finding the Steiner set for each $\{u_j,u_k,u_l\}\in N_G(z)$ incurs $O(n^5)$ time.  It is clear that the running time of Algorithm \ref{algo2} is $O(n^5)$.  Hence the following theorem holds.
	\begin{theorem}\label{tcsi}
		Let $G$ be a triad-convex split graph. A minimum Steiner set $S$ of $G$ for $R=I$ can be computed in $O(n^5)$ time, where $n$ is the number of vertices in $G$.
	\end{theorem}
	\noindent
	\textbf{Application 2: Circular-convex split graphs}
	\\\\
	We shall explore the possibility of solving STREE on circular-convex split graphs using the algorithm of STREE of path-convex split graphs as a black box.  We present a polynomial-time reduction to map the instances of circular-convex split graphs to the instances of path-convex split graphs.  The reduction is similar to the reduction presented in \cite{pandey2019domination}. 
	\\
	Let $G$ be a circular-convex split graph with $|K|=m$ and $|I|=n$.  Let the circular ordering $\prec$ on $I$, say $x_1\prec x_2 \prec \ldots \prec x_m \prec x_{m+1}=x_1$, such that for each $v\in K$, $N^I_G(v)$ is a circular arc.  For each $v\in K$, let $N^I_G(v)$ be $\{x_a,x_{a-1},\ldots,x_{b-1},x_b\}$, and $l(u)=x_a$,~$r(u)=x_b$.
	\\
	The following lemma is a key result for our reduction to work.
	\begin{lemma}\label{lci}
		Let $G$ be a circular-convex split graph.  Let $S$ be a minimum Steiner set of $G$ for $R=I$.  Then, for a vertex $z\in I$, $1\leq |N_G(z)\cap S|\leq 2$.
	\end{lemma}
	\begin{proof}
		It is clear that for every vertex $z\in R$, $|N_G(z)\cap S|\geq 1$.
		\\
		Suppose that $|N_G(z)\cap S|\geq 3$.  We observe that $N^I_G(N_G(z))$ is a circular arc in $I$ containing $z$.  Let the endpoints of that circular arc be $x_i,x_j$, for some $i,j$, $1\leq i<j\leq n$.  Then there exist two vertices $w_k,w_l\in N_G(z),~w_k,w_l\in K,~1\leq k<l\leq m$ such that $x_i\in N_G(w_k)$ and $x_j\in N_G(w_l)$.  It is clear that $N_G(w_k)\cup N_G(w_l)= N^I_G(N_G(z))$.  Consider $S'=(S\setminus N_G(z))\cup \{w_k,w_l\}$.  Note that $|S'|\leq |S|-1$, which is a contradiction that $S$ is a minimum Steiner set of $G$ for $R=I$.
		%For each $u\in N(x_i)$, $N_G(u)$ is a circular-arc in $I$ containing $x_i$.
		% $N_G(N_G(x_i))$ is also a circular arc in $I$ containing $x_i$  
		%Let $x_c$ and $x_d$ are two endpoints of this arc.  Then there exist some $w_j,w_k\in N_G(x_i)$ such that $x_c\in N^I_G(w_j)$ and $x_d\in N^I_G(w_k)$.  Then the vertices $w_j,w_k$ connects all the vertices of circular arc $N^I_G(N_G(x_i))$.  Hence, if for a minimum Steiner set $S$ of $G$ when $R=I$, $|N_G(x_i)\cap S|\geq 3$, then $S'=(S\setminus N_G(x_i))\cup \{w_j,w_k\}$ is also a minimum Steiner set of $G$ for $R=I$ and $|S'|<|S|$, which is a contradiction that $S$ is a minimum Steiner set of $G$ for $R=I$.
		\qed
	\end{proof}
	The above lemma indicates either $|N_G(z)\cap S|= 1$ or $|N_G(z)\cap S|=2$.  Accordingly, for each circular-convex split graph, we construct a corresponding path-convex split graph as part of Construction 4 and Construction 5 which are explained below.  Further, using our construction and Algorithm \ref{apci}.  We obtain a polynomial-time algorithm for circular-convex split graphs.  We choose an arbitrary vertex, say $x_i$ from $I$.  Since $x_i\in R$, we have the following cases:
	\\\\
	Case 1: Exactly one neighbor of $x_i$ is in $S$.
	\\
	Case 2: Exactly two neighbors of $x_i$ is in $S$.
	\\\\
	Let $N_G(x_i)=\{u_1,\ldots,u_s\}$.  For each $u_j\in N_G(x_i)$, $1\leq j \leq s$, $A(u_j)$ represents a minimum Steiner set of $G$ for $R=I$ containing $u_j$.  For each $u_j,u_k\in N_G(x_i)$, $1\leq j<k \leq s$, $A(u_j,u_k)$ represents a minimum Steiner set of $G$ for $R=I$ containing $u_j,u_k$.
	%\begin{lemma}
	%	The Steiner set $S$ of $G$ when $R=I$ is minimum, and $|S_1\cup S_2|=O(|I|^2)$.
	%\end{lemma}
	\\
	\noindent\textbf{Computation of minimum Steiner sets for Case 1 and Case 2.}
	\\In each of the two cases, corresponding to the circular-convex split graph $G$, we define a path-convex split graph.
	\\\textbf{Case 1: Exactly one neighbor of $x_i$ is in $S$.}
	\\\textbf{Construction 4: We define the graph $G_j$ as follows;}
	\\For each $u_j\in N_G(x_i),1\leq j\leq s$, we do the following:
	\\Let the endpoints of $N^I_G(u_j)$ be $l(u)=x_a,r(u)=x_b$.  Since $N_G(x_i)\setminus \{u_j\}$ is not in $S$, using the graph $G-(N_G(x_i))$, we construct $G_j$ as follows; $V(G_j)=(V(G)\setminus N_G[x_i])\cup \{\alpha_1,\alpha_2, \beta_1,\beta_2\}$, $K_j=(K\cap V(G_j))\cup \{\alpha_1,\alpha_2\}$, $I_j=(I\cap V(G_j))\cup \{\beta_1,\beta_2\}$, and $E(G_j)=\{\{x,y\}\mid x,y\in V(G_j), \{x,y\}\in E(G)\}\cup \{\{\alpha_1,p\},\{\alpha_2,p\}\mid p\in (K\cap V(G_j))\}\cup \{\{\alpha_1,q\},\{\alpha_2,r\}\mid q\in \{x_a,x_{a+1},\ldots,x_{i-1}\},~r\in \{x_{i+1},\ldots,x_{b-1},x_b\}\}\cup \{\{\alpha_1,\beta_1\},\{\alpha_2,\beta_2\},\{\alpha_1,\alpha_2\}\}$.
	%Split $x_i$ into $x'_i,x''_i$.  Split $w_j$ into $w'_j,w''_j$.  The set $N_G(w_j)$ is a circular arc in $I$ containing $x_i$.  Let $x_a$ and $x_b$ denote the endpoints of this arc.  Join $w'_j$ with the vertices of this arc present on one side of $x_i$.  Join $w''_j$ with the vertices of this arc present on another side of $x_i$.  Join $x'_i$ with $w'_j$ and $x''_i$ with $w''_j$.  Observe that $x'_i$ and $x''_i$ are pendant vertices, thereby ensures that $w'_j$ and $w''_j$ are in the solution.  Formally, $G_{w_j}$ with $K_{w_j}=(K\setminus N_G(x_i))\cup \{x'_i,x''_i\}$, $I_{w_j}=(I\setminus \{x_i\})\cup \{x'_i,x''_i\}$, and $E(G_{w_j})=\{\{a,b\}\in E(G)\mid a\in K\setminus N_G(x_i),~b\in I\setminus \{x_i\}\}\cup \{x'_i,w'_j\}\cup \{x''_i,w''_j\}$.  Let the minimum Steiner set obtained for $G_{w_j}$ when $R=I$ be $S_{w_j}$.
	\begin{lemma}
		$G_j$ is a path-convex split graph.
	\end{lemma}
	\begin{proof}
		We prove that $G_j$ is a path-convex split graph, by providing a linear ordering $\sigma$ on $I$.  The ordering $\sigma$ on $I$ is $\beta_2\prec x_{i+1}\ldots \prec x_b\prec \ldots \prec x_a\prec \ldots x_{i-1}\prec \beta_1$.  We can observe that for every $v\in K_j$, $N_{G_j}(v)$ is consecutive in $\sigma$.  Therefore, $G_j$ is a path-convex split graph.
		\qed
	\end{proof}
	Let $S_j$ be a minimum Steiner set $S_j$ of $G_j$ for $R=I_j$ is obtained using Algorithm \ref{apci}.  Then a minimum Steiner set of $G$ for $R=I$ containing $u_j$ is $A(u_j)=S_j\cup \{u_j\}$ and the proof for minimality of $A(u_j)$ is established in Lemma \ref{lci1}.
	\begin{lemma}\label{lci1}
		For some $u_j\in N_G(x_i)$, if $N_G(x_i)\cap S=\{u_j\}$ then $|S|=|S_j|-1$.
	\end{lemma}
	\begin{proof}
		We know that $S_j$ is a minimum Steiner set of $G_j$ for $R=I$ containing $\alpha_1,~\alpha_2$.  Construct the graph $G'_j$ from $G_j$ as follows; $V(G'_j)=(V(G_j)\setminus \{\alpha_1,\alpha_2, \beta_1,\beta_2\})\cup \{u_j,x_i\}$, and $E(G'_j)=\{\{u_j,p\}\mid p\in (N_{G_j}(\alpha_1)\cup N_{G_j}(\alpha_2))\}\cup \{\{x,y\}\mid x,y\in V(G'_j),~\{x,y\}\in E(G_j)\}\cup \{\{x_i,q\}\mid q\in (N_{G_j}(\beta_1)\cup N_{G_j}(\beta_2))\}\cup \{\{u_j,r\}\mid r\in K_j\}$.
		%Merge $w'_j,~w''_j$ to a single vertex $w_j$ and $x'_i,~x''_i$ to a single vertex $x_i$.  
		The set $(S_j\setminus \{\alpha_1,~\alpha_2\}) \cup \{u_j\}$ is a Steiner set of $G'_j$ for $R=I$.  Observe that $G'_j$ is a subgraph of $G$, and $V(G)=V(G'_j)\cup N_G(x_i)$.  Since $u_j$ connects $N_G[x_i]$, the set $(S_j\setminus \{\alpha_1,~\alpha_2\}) \cup \{u_j\}$ is also a Steiner set of $G$.  For any minimum Steiner set $S$ of $G$ containing $u_j$ for $R=I$, clearly, $|S|\leq |S_j|-1$.
		\\\\
		Suppose that $S$ is a minimum Steiner set of $G$ for $R=I$ such that $S\cap N_G(x_i)=u_j$.  Consider the graph $G'=G-(N_G(x_i)\setminus \{u_j\})$.  Observe that $S$ is a Steiner set of $G'$ for $R=I$.  Now we construct $G_j$ from $G'$ by using Construction 4.  For $G_j$ the set $(S\setminus \{u_j\})\cup \{\alpha_1,~\alpha_2\}$ is a minimum Steiner set for $R=I$.  Hence for a minimum Steiner set $S_j$ of $G_j$, $|S_j|\leq |S|+1$.
		\\
		Thus $|S|=|S_j|-1$.
		\qed
	\end{proof}
	\noindent\textbf{Case 2: Exactly two neighbors of $x_i$ is in $S$.}
	\\\textbf{Construction 5:  We define the graph $G_{jk}$ as follows;}
	\\For each $u_j,u_k\in N_G(x_i),~1\leq j<k\leq s$, we do the following:
	\\Let the endpoints of $N^I_G(u_j)\cup N^I_G(u_k)$ be $l(u)=x_a,~r(u)=x_b$.  Since $(N_G(x_i))\setminus \{u_j,u_k\}) \cap S=\emptyset$, from the graph $G-(N_G(x_i)\setminus \{u_j,u_k\})$, we construct $G_{jk}$ as follows; $V(G_{jk})=(V(G)\setminus (N_G[x_i]\setminus \{u_j,u_k\}))\cup \{\beta_1,\beta_2\}$ with $K_{jk}=(K\cap V(G_{jk}))$, $I_{jk}=(I\cap V(G_{jk}))\cup \{\beta_1,\beta_2\}$, and $E(G_{jk})=\{\{x,y\}\mid x,y\in V(G_{jk}) ,~\{x,y\}\in E(G)\} \cup \{\{u_j,p\}, \{u_k,q\}\mid p\in \{x_a,x_{a+1},\ldots,x_{i-1}\},~q\in \{x_{i+1},\ldots,x_{b-1},x_b\}\}\cup \{\{u_j,\beta_1\},\{u_k,\beta_2\}\}$.
	%remove $N(x_i)\setminus \{w_j,w_k\}$.  Now we shall convert the resultant graph into a path-convex split graph, we perform the following steps; Split $x_i$ into $x'_i,~x''_i$.  The set $N_G(\{w_j,w_k\})$ is a circular arc in $I$ containing $x_i$.  Let $x_a,x_b$ denote the endpoints of this arc.  Join $w_j$ with the vertices of this arc present on one side of $x_i$. Join $w_k$ with the vertices of this arc present on another side of $x_i$. Join $x'_i$ with $w_j$, and $x''_i$ with $w_k$.  It is clear that $x'_i$ and $x''_i$ are pendant vertices in $G_{w_j,w_k}$.
	\begin{lemma}
		$G_{jk}$ is a path-convex split graph.
	\end{lemma}
	\begin{proof}
		We prove that $G_{jk}$ is a path-convex split graph, by providing a linear ordering $\sigma$ on $I$.  The ordering $\sigma$ on $I$ is $\beta_2\prec x_{i+1}\ldots \prec x_b\prec \ldots \prec x_a\prec \ldots x_{i-1}\prec \beta_1$.  We can observe that for every $v\in K_{jk}$, $N_{G_{jk}}(v)$ is consecutive in $\sigma$.  Therefore, $G_{jk}$ is a path-convex split graph.
		\qed
	\end{proof}
	\noindent Let $S_{jk}$ be a minimum Steiner set of $G_{jk}$ for $R=I_{jk}$ is obtained using Algorithm \ref{apci}.  Then a minimum Steiner set of $G$ for $R=I$ containing $u_j,u_k$ is $A(u_j,u_k)=S_{jk}$, and the proof for minimality of $A(u_j)$ is established in Lemma \ref{lci2}.
	\begin{lemma}\label{lci2}
		For some $u_j,u_k\in N_G(x_i)$, if $N_G(x_i)\cap S=\{u_j,u_k\}$ then $|S|=|S_{jk}|$.
	\end{lemma}
	\begin{proof}
		We know that $S_{jk}$ is a minimum Steiner set of $G_{jk}$ for $R=I_{jk}$ containing $u_j,~u_k$.  Construct the graph $G'_{jk}$ from $G_{jk}$ as follows; $V(G'_{jk})=(V(G_{jk})\setminus \{\beta_1,\beta_2\})\cup \{x_i\}$, and $E(G'_{jk})=\{\{x,y\}\mid \{x,y\}\in E(G_{jk}),~x,y\in V(G'_{jk})\} \cup \{\{x_i,q\}\mid q\in (N_{G_{jk}}(\beta_1)\cup N_{G_{jk}}(\beta_2))\}$ with $I'_{jk}=(I_{jk}\setminus \{\beta_1,\beta_2\})\cup \{x_i\}$.  The set $S_{jk}$ is a Steiner set of $G'_{jk}$ for $R=I'_{jk}$.  Observe that $G'_{jk}$ is a subgraph of $G$, and $V(G)=V(G'_{jk})\cup N_G(x_i)$.  Since $u_j,~u_k$ connects $x_i$ and $N^I_{G_{jk}}(N_{G_{jk}}(x_i))$ , the set $S_{jk}$ is also a Steiner set of $G$.  For any minimum Steiner set $S$ of $G$ containing $u_j,~u_k$ for $R=I$, clearly, $|S|\leq |S_{jk}|$.
		\\\\
		Suppose that $S$ is a minimum Steiner set of $G$ for $R=I$ such that $S\cap N_G(x_i)=\{u_j,u_k\}$.  Consider the graph $G'=G-(N_G(x_i)\setminus \{u_j,u_k\})$.  Observe that $S$ a Steiner set of $G'$ for $R=I$.  Now we construct $G_{jk}$ from $G'$ by using Construction 5.  For $G_{jk}$, $S$ is a Steiner set for $R=I$.  Hence for a minimum Steiner set $S_{jk}$ of $G_{jk}$, $|S_{jk}|\leq |S|$.
		\\
		Thus $|S|=|S_{jk}|$.
		\qed
	\end{proof}
	\noindent
	We shall present an algorithm to find a minimum Steiner set of $G$ for $R=I$.
	\begin{algorithm}[H]
		\caption{{\em STREE for circular-convex split graphs.}}
		\label{acci}
		\begin{algorithmic}[1]
			\State{{\tt Input:}  A connected circular-convex split graph $G$, $R=I$. }
			\State{Let $S=\emptyset$, $S_1=\emptyset$, $S_2=\emptyset$.}
			\State{Choose an arbitrary vertex, say $x_i\in I$.}
			\ForAll{$u_j,~u_j\in N_G(x_i)$}
			\State{Construct $G_{j}$ using construction 4.}
			\State{Using Algorithm \ref{apci}, find a minimum Steiner set $S_j$ of $G_{j}$ for $R=I_j$.}
			\State{Update$S_1=S_1\cup \{S_j\}$.}
			\EndFor
			\ForAll{$u_j,u_k,~u_j,u_k\in N_G(x_i)$}
			\State{Construct $G_{jk}$ using construction 5.}
			\State{Using Algorithm \ref{apci}, find a minimum Steiner set $S_{jk}$ of $G_{jk}$ for $R=I_j$.}
			\State{Update $S_2=S_2\cup \{S_{jk}\}$.}
			\EndFor
			\State{The minimum cardinality set among $S_1\cup S_2$ is $S$.}
		\end{algorithmic}
	\end{algorithm}
	\noindent The proof of correctness of Algorithm \ref{acci} follows from Lemmas \ref{lci}, \ref{lci1}, \ref{lci2}.  Observe that in Case 1 for $u_j\in N_G(z)$, the time required for constructing $G_j$ is $O(n^2)$.  Finding the Steiner set for $G_j$ incurs $O(n^2)$ time.  Thus finding the Steiner set for each $u_j\in N_G(z)$ incurs $O(n^3)$ time.  Similarly, for Case 2, the time required for constructing $G_{jk}$ is $O(n^2)$.  For finding the Steiner set for $G_{jk}$ incurs $O(n^2)$ time.  Thus finding the Steiner set for each $u_j,u_k\in N_G(z)$ incurs $O(n^4)$ time.  It is clear that the running time of Algorithm \ref{algo2} is $O(n^4)$.  Hence the following theorem holds.
	\begin{theorem}
		Let $G$ be a circular-convex split graph.  A minimum Steiner set $S$ of $G$ for $R=I$ can be computed in $O(n^4)$ time, where $n$ is the number of vertices in $G$.
	\end{theorem}
	%Algorithm \ref{acci} for circular-convex split graph runs in time $O(mn^2)$.
	\noindent Having analyzed the P-versus-NPC status of STREE for convex split graphs with convexity on $I$, we shall now analyze the same with respect to split graphs having convexity on $K$.	
	\subsection{STREE in split graphs with convexity on $K$}\label{sck}
	When we refer to convex split graphs in this section, we refer to convex split graphs with convexity on $K$.  For STREE on split graphs with convexity on $K$, we establish hardness results for chordal-convex split graphs, and polynomial-time algorithms for tree-convex, and circular-convex split graphs.  %For tree-convex and circular-convex we show that STREE is polynomial-time solvable, whereas for chordal-convex we show that STREE is NP-complete.
	%When we refer to convex split graphs in this section, we refer to convex split graphs with convexity on $I$.
	%Since split graphs with convexity on $I$ has both polynomial and NP-complete complexities for STREE, it is natural to analyze the complexity of STREE in split graphs with convexity on $K$.  In this section, we look into tree-convex, circular-convex, and chordal-convex split graphs with convexity on $K$.  Therefore in Section \ref{sck} when we refer to convex split graphs, we refer to convex split graphs with convexity on $K$.
	\subsubsection{Tree-convex split graphs}\label{stk}
	In this section, we present a polynomial-time algorithm to find a minimum Steiner tree on tree-convex split graphs. % We solve for first the case $R=I$, and using which we solve (i) $R\subset I$ and (ii) $R\cap K\neq \emptyset$ and $R\cap I\neq \emptyset$.
	%We solve the case $R=I$, and we know that $R\subset I$ and  $R\cap K\neq \emptyset$ and $R\cap I\neq \emptyset$ can be solved by using $R=I$ approach.  
	Let $G$ be a tree-convex split graph with an imaginary tree $T$ on $K$.  We present a polynomial-time algorithm to find a Steiner set $S$ of $G$ for $R=I$.  We work with the underlying imaginary tree $T$ on $K$ to compute $S$ for $R$. 
	%\\To construct the Steiner tree for $R=I$ on $G$, we use the following scheme.  
	As part of our algorithm to compute $S$, we color vertices of $T$ (gray, white, and black).  Initially, all vertices in the imaginary tree $T$ are colored gray.  The vertex colored gray is changed to white or black as per the following rules:
	%\\We recolor the gray-colored vertex as white or black as per the following rules:
	\\\textbf{Rule 1:}(Gray-colored vertex is changed to white) The color of a leaf $u\in T$ is changed to white when there does not exist a pendant vertex in  $N^I_G(u)$.
	\\\textbf{Rule 2:}(Gray-colored vertex is changed to Black) The color of a leaf $u\in T$ is changed to black when there exists a pendant vertex in $N^I_G(u)$.
	%\\The algorithm that computes Steiner tree for the case $R=I$ works with imaginary tree $T$. 
	\\Our algorithm employs Rule 1 and Rule 2 iteratively in computing $S$.  To begin with, we choose an arbitrary leaf vertex, say $u$ in $T$.  If  Rule 1 is applicable, then $G$ is modified to $G=G-\{u\}$ and $T=T-\{u\}$.  If Rule 2 is applicable, then $S=S\cup \{u\}$ and $G$ is modified to $G=G-N_G^I[u]$, $T=T-\{u\}$.  We continue the process for $|K|-1$ times.  
	\\Observe that as per this coloring scheme gray colored vertex is recolored to white or black.  The recoloring happens exactly once for each vertex in $T$.  The vertices that are colored black during the process are included in the set $S$.  We shall now show that the set $S$ is indeed minimum in Theorem \ref{tk}.  
	%{\bf Algorithm 2: Finding a minimum Steiner set in a tree-convex split graph with convexity on $K$}\\
	%\\\emph{Step 0.} [Initialize] Set $S=\emptyset$.
	%\\\emph{Step 1.} [Delete $|K|-1$ leaf vertices one at a time] Choose a leaf vertex say $u\in T$.  Let $v$ be the vertex adjacent to $u\in T$
	%\begin{itemize}
	%	\item[] If $u$ is colored white, then remove the vertex $u$ from $G$. Set $G=G-u$.
	%	\item[] If $u$ is colored black, then $S=S\cup \{u\}$, and remove the vertex $u,~N^I_G(u)$ from $G$. Set $G=G-u$
	%	\item[] If $u$ is colored gray, and $N^I_G[u]$ does not have any pendant vertex, then change the color of $u$ to white.
	%	\item[] If $u$ is colored gray, and $N^I_G[u]$ has a pendant vertex adjacent to it, then change the color of $u$ to black.
	%\end{itemize}
	%\noindent \emph{Step 2.} [Process last vertex] If $u\in T$ and $|N^I_G(u)|\geq 1$, then $u$ is colored black, and $u$ is colored white otherwise.
	%\noindent  Due to page constraint an example is included in Appendix Figure \ref{tree1} and its trace is included in Appendix figures \ref{tree2}, \ref{tree3}, \ref{tree4}.
	%Due to page constraint an example and its trace is included in Appendix Figure \ref{path}
	%Let $\{u_1,\ldots,u_p\}$ be the vertices included by the algorithm.  Without loss of generality, we shall arrange 
	\\
	Let $G$ be a tree-convex split graph with imaginary tree $T$.  Without loss of generality, we shall arrange the vertices in $T$ (with the assumption that $T$ is a rooted tree) as per BFS order $(w_1,\ldots,w_m)$.  Let the vertices in Level $i$ be $V_i$, $1\leq i \leq k, ~k$ denotes the height of $T$ such that the root is in level 1.  Let $S=(a_1,\ldots,a_p)$ denotes vertices chosen by our algorithm.  Let $S'=(b_1,\ldots,b_q)$ denote any optimal Steiner of $G$.
	\\
	Since $S$ is an optimal Steiner set, it true that $|S'|\leq |S|$.  To show that $|S|=|S'|$, we need to prove that $|S'|\geq |S|$.  We prove $|S'|\geq |S|$ by using Theorem \ref{tk}.
	\begin{theorem}\label{tk}
		For each Level $i,~1\leq i \leq k$, $|\bigcup\limits_{i=1}^k V_i\cap S'|\geq |\bigcup\limits_{i=1}^k V_i\cap S|$.
	\end{theorem}
	\begin{proof}
		On the contrary, $|\bigcup\limits_{i=1}^k V_i\cap S'|< |\bigcup\limits_{i=1}^k V_i\cap S|$.  Let $j$ be the maximum level at which $|\bigcup\limits_{i=j}^k V_i\cap S'|< |\bigcup\limits_{i=1}^j V_i\cap S|$.  Then there exists $v\in K$ such that $v\in V_j\cap S$ and $v\notin V_j\cap S'$.  Observe that the algorithm has included $v$ because of Rule 2 and it is adjacent to a pendent vertex in $I$, say $z$ in that iteration.  since $S'$ is an optimal solution, there exists $u\in (N_G(z)\cap S')$.  It is clear that $u\notin S$, and $u\in (V_{j+1}\cup \ldots \cup V_k)$.  Without loss of generality we shall assume that $u$ is in level $r,~(j+1)\leq r \leq k$.  Since at level $r$, $u\in (V_r\cap S')$ and $u\notin (V_r\cap S)$, we continue this for each $v\in (V_j\cap S)$ and $v\notin (V_j\cap S')$, and it contradicts that $|\bigcup\limits_{i=j}^k V_i\cap S'|< |\bigcup\limits_{i=j}^k V_i\cap S|$.  We continue this argument for each level, and stop this argument when we reach level 1.  Thus we obtain $|\bigcup\limits_{i=1}^k V_i\cap S'|\geq |\bigcup\limits_{i=1}^k V_i\cap S|$.
		\qed
	\end{proof}	
	\noindent It is clear that $|S'|\geq |S|$.  Therefore, $S$ is also an optimal solution of $G$.
	\\\\
	\emph{Remarks:} Since STREE on tree-convex split graphs is polynomial-time solvable, STREE is polynomial-time solvable on well-known special structures such as path, triad, star, and comb-convex split graphs.  It is important to note that the above approach can be used as a black box for STREE on circular-convex split graphs.
	\\\\
	\textbf{Application 3:  Circular-convex split graphs}
	\\\\
	We investigate the classical complexity of STREE on circular-convex split graphs.  We shall explore the possibility of solving STREE on circular-convex split graphs using the algorithm for STREE on path-convex split graphs (subclass of tree-convex split graphs) as a black box
	We now provide a polynomial-time reduction to map the instances of circular-convex split graphs to the instances of path-convex split graphs.
	%We have proved that STREE on tree-convex split graphs is polynomial-time solvable.  Further, we focus on the split graphs having an imaginary cycle on $K$ such that each vertex in $I$ is adjacent to consecutive vertices in the imaginary cycle.  In this section, we prove that STREE on the circular-convex split graph is polynomial-time solvable.  
	Let $G$ be a circular-convex split graph with $|K|=m$ and $|I|=n$.  Let the circular ordering $\prec$ on $K$, say $w_1\prec w_2\prec \ldots \prec w_t \prec w_{t+1}=w_1$, such that for each $z\in I$, $N_G(z)$ is a circular arc.
	\\
	The following lemma is a key result for our reduction to work.
	\begin{lemma}\label{lck}
		Let $G$ be a circular-convex split graph.  Let $S$ be a minimum Steiner set of $G$ for $R=I$.  Let the minimum degree vertex in $I$ be $z$.  Then, $1\leq |N_G(z)\cap S|\leq 2$.
	\end{lemma}
	\begin{proof}
		It is clear that for every vertex $y\in R$, $|N_G(y)\cap S|\geq 1$.
		\\
		Suppose that $|N_G(z)\cap S|\geq 3$, where $z$ is the minimum degree vertex in $I$.  Observe that $N_G(z)$ is a circular arc in $K$.  Let the endpoints of circular arc obtained by $N_G(z)$ be $w_i,w_j$. It is clear that $N^I_G(w_{i+1})\cup \ldots \cup N^I_G(w_{j-1})\subseteq N^I_G(w_i)\cup N^I_G(w_j)$.  Consider $S'=(S\setminus N_G(z))\cup \{w_i,w_j\}$.  Note that $|S'|\leq |S|-1$, which is a contradiction that $S$ is a minimum Steiner set of $G$ for $R=I$.
		\qed
	\end{proof}
	The above lemma indicates either $|N_G(z)\cap S|=1$ or $|N_G(z)\cap S|=2$.  Accordingly for each circular-convex split graph, we construct a corresponding path-convex split graph as part of Construction 6 and Construction 7 which are explained below.  Further, using our construction and Algorithm \ref{apci}.  We obtain a polynomial-time algorithm for circular-convex split graphs.  We have the following cases:
	\\\\
	Case 1: Exactly one neighbor of $z$ is in $S$.
	\\
	Case 2:  Exactly two neighbors of $z$ is in $S$.
	\\\\
	Let $N_G(z)=\{u_1,\ldots,u_s\}$.  For each $u_j\in N_G(z),~1\leq j \leq s$, $A(u_j)$ represents a minimum Steiner set of $G$ for $R=I$ containing $u_j$.  For each $u_j,u_k\in N_G(z),1\leq j<k\leq s$, $A(u_j,u_k)$ represents a minimum Steiner set of $G$ for $R=I$ containing $u_j,u_k$.
	\\\textbf{Computation of minimum Steiner sets for Case 1 and Case 2.}
	\\For each case, we shall construct a path-convex split graph corresponding to the circular-convex split graph.
	\\
	\textbf{Case 1: Exactly one neighbor of $z$ is in $S$.}
	\\\textbf{Construction 6: We define the graph $G_j$ as follows;}
	\\For each $u_j\in N_G(z),~1\leq j \leq s$, we do the following:
	\\Let $N_G(z)$ be $\{w_i=u_1,\ldots,w_k=u_s\}$Since $(N_G(z)\setminus \{u_j\})\cap S=\emptyset$, we consider the graph $G_j$ with $K_j=K\setminus N_G(z)$, $I_j=I\setminus N^I_G(u_j)$, $E(G_j)=\{\{x,y\}\mid x,y\in V(G_j),~\{x,y\}\in E(G)\}$.
	\begin{lemma}
		$G_j$ is a path-convex split graph.
	\end{lemma}
	\begin{proof}
		We prove that $G_j$ is a path-convex split graph, by providing a linear ordering $\sigma$ on $K_j$.  Let the ordering be $w_{i+1}\prec w_{i+2}\prec \ldots \prec w_{k+1}$.  We can observe that for every $y\in I$, $N_{G_j}(y)$ is consecutive in $\sigma$.  Therefore $G_j$ is a path-convex split graph.
		\qed
	\end{proof}
	Let $S_j$ be a minimum Steiner set of $G_j$ for $R=I_j$ is obtained using the algorithm for STREE on path-convex split graphs (subclass of tree-convex split graphs).
	\begin{lemma}\label{lck1}
		If $N_G(z)\cap S=\{u_j\}$, then $|S|=|S_j|+1$.
	\end{lemma}
	\begin{proof}
		Suppose that $S_j$ is a minimum Steiner set of $G_j$ for $R=I$.  Construct $G'_j$ as follows; $V(G'_j)=V(G_j)\cup \{u_j\}\cup \{z\}$ and $E(G'_j)=E(G_j)\cup \{\{u_j,z\}\}$ with $I'_j=I_j\cup \{z\}$.  Then $S'_j=S\cup \{u_j\}$ is a Steiner set of $G'_j$ for $R=I'_j$.  Observe that $G'_j$ is a subgraph of $G$, and $V(G)=V(G'_j)\cup N^I_G(u_j)\cup N_G(z)$.  Hence $S=S'_j$ is also a Steiner set of $G$ for $R=I$.  For any minimum Steiner set $S$ of $G$ for $R=I$ containing $u_j$, clearly, $|S|\leq |S_j|+1$.
		\\
		Suppose that $S$ is a minimum Steiner set of $G$ for $R=I$ such that $S\cap N_G(z)=\{u_j\}$.  Consider the graph $G'=G-(N_G(z)\cup N^I_G(u_j))$ with $I'_j=I\setminus \{z\}$.  The set $S\setminus \{u_j\}$ is a Steiner set of $G'$ for $R=I'_j$.  The resultant graph is $G_j$.  Thus $S_j=S\setminus \{u_j\}$ is a minimum Steiner set of $G_j$ for $R=I$.  Hence for a minimum Steiner set $S_j$ of $G_j$, $|S_j|\leq |S|-1$.
		\\Therefore, $|S|=|S_j|+1$.
		\qed
	\end{proof}
	\textbf{Case 2:  Exactly two neighbors of $z$ is in $S$.}
	\\\textbf{Construction 7:  We define the graph $G'$ as follows;}
	\\For each $u_j,u_k\in N_G(z)$, $1\leq j<k\leq s$, we do the following:
	\\Since $(N_G(z)\setminus \{u_j,u_k\})\cap S=\emptyset$, from the graph $G-(N_G(z)\setminus \{u_j,u_k\})$, we construct $G_{jk}$ as follows; $V(G_{jk})=K_{jk}\cup I_{jk}$, $K_{jk}=K\setminus \{u_j,u_{j-1},\ldots,u_k\}$, $I_{jk}=I\setminus (N^I_G(u_j)\cup N^I_G(u_k))$, $E(G_{jk})=\{\{x,y\}\mid x,y\in V(G_{jk}),~\{x,y\}\in E(G)\}$. 
	\begin{lemma}
		$G_{jk}$ is a path-convex split graph.
	\end{lemma}
	\begin{proof}
		We prove that $G_{jk}$ is a path-convex split graph, by providing a linear ordering $\sigma$ on $I$.  The ordering $\sigma$ on $I$ is $u_{j+1}\prec u_{j+2}\prec \ldots u_{k+1}$.  We can observe that for every $v\in K_{jk}$, $N_{G_{jk}}(v)$ is consecutive in $\sigma$.  Therefore, $G_{jk}$ is path-convex split graph.
		\qed
	\end{proof}
	Let $S_{jk}$ be a minimum Steiner set $S_{jk}$ of $G_{jk}$ for $R=I_{jk}$ is obtained using Algorithm \ref{apci}.
	\begin{lemma}\label{lck2}
		If $N_G(z)\cap S=\{u_j,u_k\}$, then $|S|=|S_{jk}|+2$.
	\end{lemma}
	\begin{proof}
		We know that $S_{jk}$ is a minimum Steiner set of $G_{jk}$ for $R=I$.  Construct the graph $G'_{jk}$ from $G_{jk}$ as follows; $V(G'_{jk})=(V(G_{jk})\cup \{u_j,u_k\}$, and $E(G'_{jk})=\{\{x,y\}\mid \{x,y\}\in E(G_{jk}),~x,y\in V(G'_{jk})\} \cup \{\{z,u_j\}, \{z,u_k\}\}$.  The set $S'_{jk}=S_{jk}\cup \{u_j,u_k\}$ is a Steiner set of $G'_{jk}$ for $R=I'_{jk}$.  Observe that $G'_{jk}$ is a subgraph of $G$, and $V(G)=V(G'_{jk})\cup N_G(z)$.  Since $u_j,~u_k$ connects $N_G(z)$, the set $S'_{jk}$ is also a Steiner set of $G$.  For any minimum Steiner set $S$ of $G$ containing $u_j,~u_k$ for $R=I$, clearly, $|S|\leq |S_{jk}|$.
		\\
		Suppose that $S$ is a minimum Steiner set of $G$ for $R=I$ such that $S\cap N_G(z)=\{u_j,u_k\}$.  Consider the graph $G'=G-(N_G(z)\setminus \{u_j,u_k\})$.  Observe that $S$ a Steiner set of $G'$ for $R=I$.  Now we construct $G_{jk}$ from $G'$ by using Construction 5.  For $G_{jk}$, $S$ is a Steiner set for $R=I$.  Hence for a minimum Steiner set $S_{jk}$ of $G_{jk}$, $|S_{jk}|\leq |S|$.
		\\
		Thus $|S|=|S_{jk}|$ 
		\qed
	\end{proof}
	\noindent
	Now we shall present an algorithm for finding the minimum Steiner set of $G$ for $R=I$.
	\begin{algorithm}[H]
		\caption{{\em STREE for circular-convex split graphs.}}
		\label{acck}
		\begin{algorithmic}[1]
			\State{{\tt Input:}  A connected circular-convex split graph $G$ with $R=I$. }
			\State{Let $S=\emptyset$, $S_1=\emptyset$, $S_2=\emptyset$.}
			\State{Choose the minimum degree vertex, say $z\in I$.}
			\ForAll{$u_j,~u_j\in N_G(z)$}
			\State{Construct $G_j$ using Construction 6.}
			\State{Using Algorithm \ref{apci}, find a minimum Steiner set $S_j$ of $G_j$ for $R=I_j$.}
			\State{Update $S_1=S_1\cup \{S_j\}$.}
			\EndFor
			\ForAll{$u_j,u_k,~u_j,u_k\in N_G(z)$}
			\State{Construct $G_{jk}$ using Construction 7.}
			\State{Using Algorithm \ref{apci}, find a minimum Steiner set $S_{jk}$ of $G_{jk}$ for $R=I_{jk}$.}
			\State{Update $S_2=S_2\cup \{S_{jk}\}$.}
			\EndFor
			\State{The minimum cardinality set among $S_1\cup S_2$ is $S$.}
		\end{algorithmic}
	\end{algorithm}
	\noindent The proof of correctness of Algorithm \ref{acck} follows from Lemmas \ref{lck1}, \ref{lck2}.  Observe that in Case 1 for $u_j\in N_G(z)$, the time required for constructing $G_j$ is $O(n^2)$.  Finding the Steiner set for $G_j$ incurs $O(n^2)$ time.  Thus finding the Steiner set for each $u_j\in N_G(z)$ incurs $O(n^3)$ time.  similarly, the time required for constructing $G_{jk}$ is $O(n^2)$.  For finding the Steiner set for $G_{jk}$ incurs $O(n^2)$ time.  Thus finding the Steiner set for each $u_j,u_k\in N_G(z)$ incurs $O(n^4)$ time.  It is clear that the running time of Algorithm \ref{algo2} is $O(n^4)$.  Hence the following theorem holds.
	\begin{theorem}
		Let $G$ be a circular-convex split graph.  A minimum Steiner set $S$ of $G$ for $R=I$ can be computed in $O(n^4)$ time, where $n$ is the number of vertices in $G$.
	\end{theorem}
	\noindent
	It turns out that STREE is Polynomial-time solvable on tree-convex and circular-convex split graphs.  Further, we explore the classical complexity of STREE on a convex split graph having a chordal graph as its imaginary structure in the next section.
	\subsubsection{Chordal-convex split graph}
	We have seen that STREE on tree-convex split graphs is polynomial-time solvable.  It is natural to ask; "Is there a property $\pi$ such that STREE on $\pi$-convex split graphs is NP-complete?"  We consider one such property namely, chordality, and show that STREE on chordal-convex split graphs is NP-complete. 
	\begin{definition}
		A split graph $G$ is called {\em chordal-convex split graph with convexity on $K$}, if there is an associated chordal graph $G'$ defined on $K$, such that for each $v\in I$, $N_G(v)$ induces a subchordal graph in $G'$.
	\end{definition}
	\begin{theorem}
		STREE on a chordal-convex split graph with convexity on $K$ is NP-complete.
	\end{theorem}
	\begin{proof}
		\textbf{STREE is in NP} Given an input instance $(G,R,k)$ of STREE, and a certificate set $S\subseteq V(G)$. By using graph traversal algorithms the connectedness of the graph induced on $R\cup S$ can be verified in deterministic polynomial time.
		\\\textbf{STREE is NP-Hard}  It is known \cite{white1985steiner}, that STREE on split graphs is NP-complete and this can be reduced in polynomial time to STREE on the chordal-convex split graph using the following reduction.  We map an instance $(G,R,k)$ of STREE on split graphs to the corresponding instance $(G^*,R^*,k'=k)$ of STREE on the chordal-convex split graph as follows:
		\\$V(G^*)=V_1\cup V_2\cup V_3$,
		\\$V_1=\{w_i\mid v_i\in V(G)\cap K\}$,
		\\$V_2=\{y_i\mid v_i\in V(G)\cap I\}$,
		\\$V_3=\{x_i\mid v_i\in V(G)\cap I\}$.
		\\We shall now describe the edges of $G^*$,
		\\$E(G^*)=E_1\cup E_2$,
		\\$E_1=\{\{w_i,w_j\},\{w_i,y_k\},\{y_k,y_l\}\mid 1\leq i<j \leq m, 1\leq k<l\leq n\}$,
		\\$E_2=\{\{w_i,x_j\}, \{x_j,y_j\}\mid \{w_i,x_j\}\in E(G),~w_i,y_j\in K^*,~x_j\in I^*\}$.
		\\We define $K=V_1\cup V_2$, $I=V_3$, and an imaginary split graph $G'$ with clique $K'$ and independent set $I'$ on $K^*$ is defined with split graph $G$ as an imaginary structure.
		\\An example is illustrated in Figure \ref{chordal}, the STREE in split graph with $k=2$ is mapped to STREE on chordal-convex split graph with $R=\{x_i\mid x_i\in I\}$, $k'=2$.
		\begin{figure}[H]
			\begin{center}
				\includegraphics[scale=0.7]{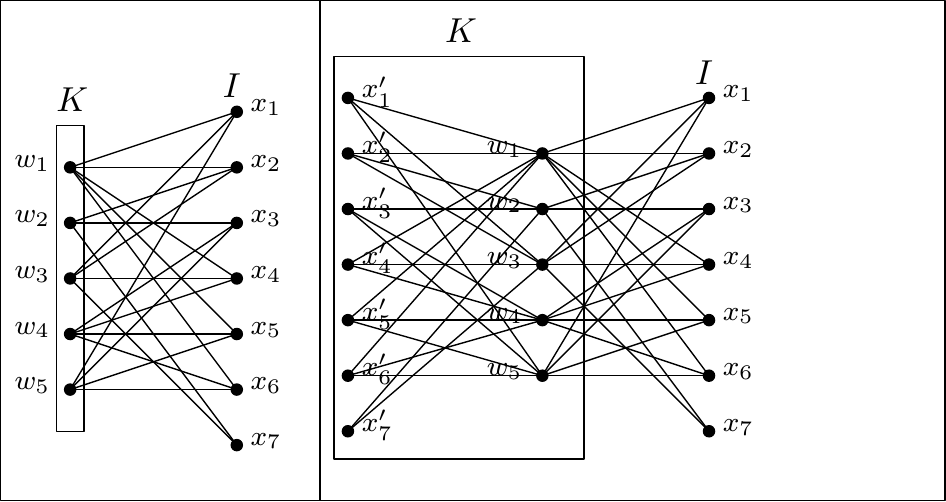}
				\caption{An example: STREE in split graphs reduces to STREE on the chordal-convex split graph.}\label{chordal}
			\end{center}
		\end{figure}
		%\noindent Since $K^*$ is a clique and we can observe from Figure \ref{chordal}, that there exists an imaginary split graph on $K^*$.  We know that split graphs are a subclass of chordal graphs, hence the graph $G^*$ is a chordal-convex split graph with convexity on $K^*$.
		\begin{myclaim}
			$G^*$ is a chordal-convex split graph.
		\end{myclaim}
		\begin{proof}
			For each $x_i\in V_3$,  $N_G(x_i)\subseteq K^*$, $K^*=V_1\cup V_2$.  By our construction, $x_i$ is adjacent to a subset of vertices in $K'$.  Therefore, the graph induced on $N_G(x_i)$ is a split graph in $G'$.  Hence $G^*$ is a chordal-convex split graph.
			\qed
		\end{proof}
		\begin{myclaim}
			STREE $(G,R,k)$ if and only if STREE $(G^*,R^*,k')$.
		\end{myclaim}
		\begin{proof}
			Since $N_G(I)\setminus N_G(I^*)=\emptyset$ and $N_G(I^*)\setminus N_G(I)=\emptyset$, $S$ is the Steiner set of $G$ is also the Steiner set of $G^*$.  Similarly, $S^*$ is the Steiner set of $G$.
			%Let $S=\{w_i~|~1\leq i \leq k\}$ is a Steiner set of size $k$ in $G$.  Then we construct the Steiner set $S^*$ of $G^*$ for $R^*=I^*$ as follows: $S^*=\{w_i~|~1\leq i\leq k,w_i\in K^*,w_i\in K\}$.  Since $S$ is a Steiner set of
			\qed
		\end{proof}
		Thus we conclude STREE is NP-Hard on the chordal-convex split graph.  Therefore, STREE is NP-complete on chordal-convex split graphs.
		\qed
	\end{proof}
	Note that yet another natural P-versus-NPC line we can observe from this paper is that the tree-convex split graph with convexity on $I$ is NP-complete whereas the tree-convex split graph is polynomial-time solvable.
	\\\\
	\section{Application 4: Domination and its variants}\label{DS}
	In this section, we prove that the solution to DS, TDS, and CDS, can be obtained from STREE of a convex split graph for $R=I$.  Let $G$ be some convex split graph.  It is known \cite{renjith2020steiner}, that a minimum Steiner set is also a minimum dominating set on the class of split graphs.  Hence the result is true for a subclass of split graphs as well.  We also show that for a split graph $G$ if dominating set $S\subseteq K$, then $S$ is also a Steiner set of $G$ for $R=I$, which we prove in the following claim.\\
	%\begin{mclaim}\label{dstree}
	%	$S$ is a minimum dominating set.
	%\end{mclaim}
	%\begin{proof}
	%	Let $D$ be a minimum dominating set of $G$. We know that $D\subseteq K\cup I$. Suppose that $D\cap I\neq \emptyset$, then let $D\cap I=\{x_1,x_2,\ldots,x_l\}$.  Including any one neighbor for each of $x_1,x_2,\ldots,x_l$ in $D$ and removing $\{x_1,x_2,\ldots,x_l\}$ from $D$ is also a minimum dominating set of $G$.  Clearly, $|S|=|D|$.  Hence $S$ is also a minimum dominating set.
	%	\qed
	%\end{proof}
	\begin{mclaim}\label{dstree}
		$S\subseteq K$ is a minimum dominating set if and only if for the Steiner tree problem when $R=I$, $S\subseteq K$ is the Steiner solution.
	\end{mclaim}
	\begin{proof}
		Let $S$ be a minimum dominating set of $G$. We know that $S\subseteq K\cup I$. Suppose that $S\cap I\neq \emptyset$, then let $S\cap I=\{x_1,x_2,\ldots,x_l\}$.  Including any one neighbor for each of $x_1,x_2,\ldots,x_l$ in $S$ and removing $\{x_1,x_2,\ldots,x_l\}$ from $S$ is also a minimum dominating set of $G$.  Hence $S$ is also a minimum Steiner set for $R=I$.
		\\
		Conversely, For the Steiner tree problem Since $S$ is a minimum Steiner set for $R=I$,  it is clear that $S$ dominates all of $I$. The vertices in $K\setminus S$ are dominated by vertices in $S$. Hence $S$ is also a minimum dominating set.
		\\Therefore, $S\subseteq K$ is a minimum dominating set if and only if for the Steiner tree problem when $R=I$, $S\subseteq K$ is the Steiner solution.
		\qed
	\end{proof}
	Since $S\subseteq K$,  by the definition of CDS, and TDS, $S$ is also a connected dominating set and total dominating set. The minimality of CDS and TDS can be proved similar to Claim \ref{dstree}.  Therefore, the following results hold;
	\begin{itemize}
		\item DS for star-convex split graphs with convexity on $I$ and comb-convex split graphs with convexity on $I$ is NP-complete.  Hence DS is NP-complete for a tree-convex split graph with convexity on $I$.
		\item DS for path-convex and triad-convex split graphs with convexity on $I$ is polynomial-time solvable.
		\item DS for a tree-convex split graph with convexity on $K$ is polynomial-time solvable.
		\item DS for a circular-convex split graph with convexity on $K$ is polynomial-time solvable.
		\item DS for a chordal-convex split graph with convexity on $K$ is NP-complete.
	\end{itemize}
	\noindent The P-versus-NPC status of STREE for split graphs with convex properties discussed in this paper also holds for DS, CDS, and TDS.
	\section{Parameterized results}\label{pc}
	In this section, we analyze the parameterized complexity of the Steiner tree problem on split graphs.  We wish to identify the tractability vs intractability status of the Steiner tree problem on split graphs.
	\\  
	In this section, we ask the following two questions and answer them;
	%We also establish the $2-\frac{1}{|I|}$ approximation algorithm for the dominating set problem on split graphs in this section.
	%\\An interesting parameterized complexity question is
	\\
	\\\emph{Whether the parameterized version of Steiner tree problem is tractable or intractable for split graphs?}
	\\We answer this question by proving that the parameterized version of Steiner tree problem with solution size as the parameter for split graphs, is W[2]-hard.  The parameterized version of Steiner tree problem which we considered with solution size being parameter $k$ (PSTREE3) is defined below:
	\begin{center}
		\fbox{\parbox[c][][c]{0.95\textwidth}{    
				\emph{PSTREE3 $(G,R,k)$}
				\\\textbf{Instance:} A split graph $G$, a terminal set $R=I$.	
				\\\textbf{Parameter:} A positive integer $k$.
				\\\textbf{Question:} Is there a set $S\subseteq V(G)\setminus R$ such that $|S|\leq k$, and $G[S\cup R]$ is connected ?
		} }
	\end{center}
	\subsection{W-hardness of STREE on split graphs}
	In this section, we show that PSTREE3 on split graphs is in W[2]-hard.  We know that the dominating set problem on split graphs is known to be W[2]-hard \cite{raman2008short}.  We give a polynomial-time reduction from the parameterized version of dominating set problem for split graphs to PSTREE2.
	\begin{theorem}
		For split graphs, PSTREE3 is W[2]-hard.
	\end{theorem}
	\begin{proof}
		We prove this by giving a reduction from the parameterized version of dominating set problem on split graphs.  We map an instance $(G,k)$ of the parameterized version of dominating set problem on split graphs to the corresponding instance $(G,R,k)$ of PSTREE2.
		\\
		We show that $G$ has a dominating set of size $k$ if and only if $G$ for $R=I$ has a Steiner set of size $k$.
		\\
		\emph{Only if:} Let $D$ be a dominating set of size $k$ in $G$.  If $D\cap I\neq \emptyset$, then by using Claim \ref{dstree}, we obtain $S$ whose cardinality is equal to the cardinality of $D$.  Clearly, $|S|=k$.
		\\
		\emph{if:} From \cite{renjith2020steiner}, it is known that $S$ is a dominating set for $G$.
		\\Therefore, PSTREE2 is W[2]-hard.
		\qed
	\end{proof}	
	\noindent Further we ask;
	\\\\\emph{Does there exists a parameter for which the corresponding parameterized version of the Steiner tree problem on split graphs is in FPT?}
	\\We prove that for the parameters such as (i)the treewidth, and (ii) the solution size and the maximum degree of $I$, then the parameterized version of Steiner tree problem for split graphs is in FPT. 
	\\
	The parameterized version of Steiner tree problem with parameter the treewidth $r$ of $G$ (PSTREE4) is defined below:
	\begin{center}
		\fbox{\parbox[c][][c]{0.95\textwidth}{    
				\emph{PSTREE3 $(G,R,k)$}
				\\\textbf{Instance:} A split graph $G$, a terminal set $R=I$.	
				\\\textbf{Parameter:} The treewidth $r$ of $G$.
				\\\textbf{Question:} Is there a set $S\subseteq V(G)\setminus R$ such that $|S|\leq k$, and $G[S\cup R]$ is connected ?
		} }	
	\end{center}
	\noindent The parameterized version of Steiner tree problem with parameters $k$ and $d$ (PSTREE5) is defined below:
	\begin{center}
		\fbox{\parbox[c][][c]{0.95\textwidth}{    
				\emph{PSTREE5 $(G,R,k)$}
				\\\textbf{Instance:} A split graph $G$, a terminal set $R=I$.	
				\\\textbf{Parameter:} A positive integer $k$, and the maximum degree $d$ of $I$.
				\\\textbf{Question:} Is there a set $S\subseteq V(G)\setminus R$ such that $|S|\leq k$, and $G[S\cup R]$ is connected ?
		} }	
	\end{center}
	\noindent We show that PSTREE4, PSTREE5 are FPT in the following section.
	\subsection{FPT algorithms for the parameterized version of the Steiner tree problem on split graphs}
	\subsubsection{FPT algorithm for PSTREE4 on split graphs with treewidth as the parameter}
	In this section, we show that PSTREE4 on split graphs exhibits an FPT algorithm when the parameters are the treewidth and the solution size.
	It is known that STREE can be solved in $3^{|R|}n^{O(1)}$ on general graphs.  We show that PSTREE4 on split graphs can be solved in $2^{|K|}n^{O(1)}$.\\
	We use the bounded search tree technique for solving PSTREE4.  We shall describe our branching algorithm; 
	Given an instance $(G,R,r)$,  we recursively branch by two cases by considering $v\in K$ is in $S$ or not in $S$.  At any iteration, if $N^I_K(v)$ contains a pendant vertex, then the branch is for one case $v\in K$ is in $S$.  In the branch where $v\in S$, we delete $v,~N^I_G(v)$ from $G$ and reduce the parameter by 1.  In the second branch, we delete $v$ from $G$ and the parameter remains the same.
	Suppose that $d^I_G(v)= 1$, then the second branch is not possible.%STOP?
	\begin{lemma}
		The solution set $S$ obtained from the above strategy is a minimum Steiner set of $G$ for $R=I$.
	\end{lemma}
	\begin{proof}
		Since $R=I$, $S\subseteq K$.  Let $|I|=m$.  By our approach, we choose an arbitrary vertex in $v\in K$ such that $d^I_G(v)\neq 0$, and we branch by having $v\in S$ and another branch with $u\notin S$.  We can observe that length of the tree is $|K|$ and the number of leaves is at most $2^{|K|}$.  For each vertex $u\in K$, we explore the two possibilities, hence one of the paths from the root to the leaf is having minimum Steiner solution.
		\qed
	\end{proof}
	Observe that by this approach we list all feasible solutions.
	Note that the length of the tree is $|K|$ and the number of leaves is at most $2^{|K|}$.  The running time of the algorithm is bounded by the number of nodes ($2^{|K|}$) and the time is taken at each node $n^c$, where $c$ is a constant.  %Hence we obtain the recurrence relation which denotes the running time of the algorithm with respect to $n=|K|$ and the solution size $k$,
	%$$T(n,k)\leq T(n-1,k-1)+T(n-1,k)+O(1), T(k,k)=O(1),T(n,0)=O(1)$$
	The algorithm runs in time $2^{|K|}n^{O(1)}$.  
	\subsubsection{FPT algorithm for PSTREE5 on split graphs with the maximum degree of $I$ and $|S|$ as the parameter}
	In this section, we show that PSTREE5 admits a kernel of size $(2d-1)k^{(d-1)}+k$.  It is known \cite{ABUKHZAM2010524}, that $d$-hitting set guarantees a kernel whose order does not exceed $(2d-1)k^{d-1}+k$.  The parameterized version of $d$-hitting set can be stated as follows:
	\begin{center}
		\fbox{\parbox[c][][c]{0.95\textwidth}{    
				\emph{d-Hitting set $(\mathcal{C},P,k)$}
				\\\textbf{Instance:} A collection $\mathcal{C}$ of subsets of size $d$ obtained from a set $P$.
				\\\textbf{Parameter:} A positive integer $k$, cardinality of every element in $\mathcal{C}$ is $d$ .
				\\\textbf{Question:} Does $\mathcal{C}$ have a hitting set of size $k$ or less ?
		} }	
	\end{center}
	We show that PSTREE 5 is FPT by using the FPT algorithm of $d$-hitting set as a black box.  Let $d=max(d_G(x_1),\ldots d_G(x_m))$ be the maximum degree among vertices in $I=\{x_1,x_2,\ldots,x_m\}$.  We convert a given split graph $G$ into a split graph $G'$ with $d$ as the degree of every vertex $z\in I$ as follows;
	\\Let $Y=y_1,y_2,\ldots,y_k$ be the vertices in $I$ whose degree is less than $d$.  Then $V(G')=V(G)\cup U,~ U=\{u_{i1},u_{i2},\ldots,u_{i(d-d_G(y_i))}\mid y_i\in Y,~1\leq i\leq k\}, E(G')=\{\{x,y\}\mid x,y\in V(G'),~\{x,y\}\in E(G)\}\cup \{\{y_i,u_{ij}\}\mid y_i\in Y,~u_{ij}\in U,~1\leq i \leq k,~1\leq j \leq d-d_G(y_i)\}\cup \{\{x,y\}\mid x\in (V(G')\cap K), ~u\in U\}$.  Observe that $K'=K\cup U$, and $I'=I$.
	\noindent\\\\
	%Let $G$ be a split graph.  We construct the corresponding split graph $G'$ such that if there exists a vertex $z\in I$ whose degree is less than $d$, then we add $d-d_G(x)$ vertices in $G'$.  For each vertex in $z\in I$ whose degree is less than $d$, we add vertices and edges such that $\{\{a_i,w_j\}, \{a_i,a_k\},\{z,a_i\}\mid v_j\in K, 1\leq j \leq p,~ i\neq k,~1\leq i \leq d-d_G(x),~1\leq k \leq d-d_G(x)\}$.  Let the vertices in $G'-G$ be $A$.
	We transform an instance of a split graph $G'$ with each $z\in I$ of degree $d$ into the corresponding instance of $d$-hitting set with $\mathcal{C}$ as a collection of subsets of a set $P$ as follows:
	\\
	The set $P=\{w_i\mid w_i\in K'\}$, and collection $\mathcal{C}=\{A_i\mid A_i=N_{G'}(x_i),~x_i\in I'\}$.  Since for each $z\in I$, have degree $d$, for each $A_i\in \mathcal{C},~1\leq i \leq m$, it is clear that cardinality of $A_i$ is $d$.  
	\\It is known that $d$-hitting set admits a kernel of size $(2d-1)k^{d-1}+k$, by using the following lemma we prove that $(G',R',k)$ admits a kernel of size $(2d-1)k^{d-1}+k$.
	\begin{lemma} \label{fpt}
		If $(\mathcal{C},P,k)$ admits a kernel of size $(2d-1)k^{d-1}+k$, then $(G',R'=I,k)$ admits a kernel of size $(2d-1)k^{d-1}+k$.
	\end{lemma}
	\begin{proof}
		Since we have a kernel for $(\mathcal{C},P,k)$, we construct a kernel for $(G',R'=I',k)$ as follows;
		For each element $\mathcal{C}$ in the crown reduction, we replace it with the corresponding $N_{G'}(x_i)$, and for each element $P$ in the crown reduction, we replace it with the corresponding vertex $w_j\in K$.  This if $(\mathcal{C},P,k)$ admits a kernel of size $(2d-1)k^{d-1}+k$, then we can transform the kernel such that $(G',R'=I,k)$ admits a kernel of size $(2d-1)k^{d-1}+k$.
		%It is true because of our transformation.
		%Let $\mathcal{C'}$ be the solution of $d$-hitting set.  By the definition of $d$-hitting set, we know that for each $A_i\in \mathcal{C}$, $A_i\cap \mathcal{C'}\neq \emptyset$.  We transform 
		%The Steiner set $S'$ contains vertices corresponding to each $w_i\in \mathcal{C'}$.  
		%Hence in $G$, for each $x_i\in I$, $S\cap N_G(x_i)\neq \emptyset$, and $S'$ is the Steiner set.  Suppose that there exists $S'$ such that $|S'|<|S|$.  Then it contradicts that $\mathcal{C'}$ is the minimum cardinality solution of $d$-hitting set.  Therefore, if $\mathcal{C'}$ is the minimum cardinality solution of $d$-hitting set, then the corresponding solution $S$ of PSTREE4 is also the minimum cardinality Steiner set.
		\qed
	\end{proof}
	\begin{theorem}
		There is a polynomial-time algorithm that, for an arbitrary instance $(G,R,k)$ of PSTREE5, either determines that it is a no instance or computes a kernel instance whose order is bounded above by $(2d-1)k^{d-1}+k$.
	\end{theorem}
	\begin{proof}
		Theorem holds because of \cite{ABUKHZAM2010524} and Lemma \ref{fpt}.
		\qed
	\end{proof}
	\noindent From Theorem \ref{fpt}, it is clear that we can obtain solution to $(G',R,k)$ in time $O(2^{(2d-1)k^{d-1}+k}n^c)$, where $c$ is a constant (by using brute-force approach for the kernel).
	\noindent  We do a polynomial-time transform from the solution $S'$ of $G'$ for $R=I'$ to the solution $S$ of $G$ for $R=I$ as follows;
	\\If $S'\cap U=\emptyset$, then $S=S'$ is the Steiner of $G$ for $R=I$.  Suppose that $S'\cap U\neq \emptyset$.  Let $\{a_1,\ldots,a_q\},~q\geq 1$ be the vertices in $S'\cap U$. 
	By the construction of $G'$, we know that $d^I_G(a_i)=1$, $a_i\in (S'\cap U)$.  
	Assume that $a_1,\ldots,a_q$ is included in $S'$ in order to connect $y_1,\ldots,y_q$.  We know that $d_{G'}(y_i)\geq 2$ in $G'$, $1\leq i \leq q$, $y_i$ is adjacent to at least one vertex in $\{w_1,\ldots,w_n\}\cap K$.  Assume without loss of generality that $y_i$ is adjacent to $w_i,~1\leq i \leq q\leq n$.  We construct $S$ of $G$ for $R=I$ as follows; $S=(S'\cap V(G))\cup \{w_i\mid \{w_i,y_i\}\in E(G)\land (N_{G'}(y_i)\cap (S'\cap U)\neq \emptyset)\}$.  Observe that $S\subseteq K$ and for each $z\in R$, $S\cap N_G(z)\neq \emptyset$.  Thus $S$ is a Steiner set of split graph $G$ for $R=I$.  Therefore, $(G,R,k)$ can be solved in time $O(2^{(2d-1)k^{d-1}+k}n^c)$, where $c$ is a constant.
	\section{Approximation algorithm for Domination on split graphs}\label{approx}
	From \cite{vazirani2001approximation}, the dominating set problem has $\log n$ approximation algorithm on general graphs, and it does not admit $(1-\epsilon)\log n$-approximation algorithm in polynomial time on general graphs, for any $\epsilon>0$, unless NP $\subseteq$ DTIME $(n^{O(\log \log n)})$.  Further, it is known \cite{BERTOSSI198437}, that the dominating set problem is NP-complete on split graphs, and we are interested in analyzing the approximation algorithm for the dominating set problem on split graphs.  In this section, we show that the dominating set problem has a $2-\frac{1}{|I|}$ approximation algorithm in polynomial time for split graphs.  Further, it is important to highlight that in \cite{CHLEBIK20081264}, it incorrectly claimed that split graphs does not admit $(1-\epsilon)\log n$-approximation algorithm in polynomial time, for any $\epsilon>0$, unless NP $\subseteq$ DTIME $(n^{O(\log \log n)})$.
	\begin{lemma}
		The dominating set problem has $2-\frac{1}{|I|}$ approximation algorithm in polynomial time for split graphs.
	\end{lemma}
	\begin{proof}
		Since the Steiner set $S$ obtained for STREE of $G$ for $R=I$ is a subset of $K$, $S$ is also a dominating set.  We know that STREE has $2-\frac{1}{|R|}$ approximation algorithm in polynomial time, where $R=I$ for split graphs.  Hence we also have $2-\frac{1}{|I|}$ approximation algorithm in polynomial time for split graphs.
		\qed
	\end{proof}
	\section{Other cases of STREE}\label{ocases}
	\noindent Having seen STREE of $G$ for $R=I$, we now consider STREE of $G$ for other cases of $R$.  Interestingly for all other cases, the solution can be obtained using the solution of STREE of $G$ for $R=I$ as a black box.
	\\\emph{Case 1:} $R=K$ or $R\subset K$.
	\\Observe that $G[R]$ is connected.  Therefore, Steiner set $S$ is an empty set.
	\\\emph{Case 2:} $R\subset I$.
	\\For $R\subset I$, we transform the graph $G$ to $G'$; $V(G')$ with $K'=K$, $I'=I\cap R$, $E(G')=\{\{u,v\}\mid u,v\in V(G'),\{u,v\}\in E(G)\}$, and $R'=I'$.  Observe that $R'\setminus R=\emptyset$ and $R\setminus R'=\emptyset$.  Thus, the solution of $(G',R')$ is precisely the solution to $(G,R)$.
	%The solution to $(G,R)$ is same as the solution of $(G',R)$.  
	%The transformation is defined as follows; $G'=G-I'$, where $I'=I\setminus R$ and $R'=R$.
	\\\emph{Case 3:} $R\cap K\neq \emptyset$ and $R\cap I\neq \emptyset$.
	\\Similar to Case 2, we obtain the solution for this case using the following transformation. Let $W=R\cap K$, and let $X=I\cap R$.  Let $G'$ be the transformed graph $V(G')$ with $K'=K\setminus W$ and $I'=X\setminus (N^I_G(W))$, $E(G')=\{\{u,v\}\mid u,v\in V(G'),\{u,v\}\in E(G)\}$, and $R'=X$.  We map the solution of $(G',R')$ to the solution of $(G,R)$ as $S=S' \cup W$.  Observe that $(S)\cap N_G(z)\neq \emptyset, ~z\in I$.  Thus for $(G,R)$, $S$ is the Steiner set.
	\\\\
	\noindent\textbf{Conclusions and directions for further research:}
	\\We have proved the classical complexity of STREE, and domination and its variants on tree-convex and circular-convex split graphs.  The results presented in this paper can be used as a framework for the Steiner tree variants (Steiner path and cycle) and the domination problems (outer-connected domination, Roman domination) restricted to split, and bipartite graphs.  
	\\We have given a $2-\frac{1}{|I|}$approximation algorithm for DS on split graphs, and it would be interesting to explore whether $c-\frac{1}{|I|}$-approximation algorithm, $1<c<2$ is possible for STREE and DS on split graphs.  
	\\We proved that the parameterized version of Steiner tree problem on split graphs with parameter being solution is W[2]-hard, and with respect to the parameters such as (i) the treewidth and the solution size, and (ii) the maximum degree of $I$ and the solution size, we have shown that their corresponding parameterized version of the Steiner tree problem is FPT.  One can look into other parameters of the Steiner tree problem on a split graph and analyze their parameterized complexity.
	\\Furthermore, one can analyze the classical complexity for STREE and DS restricted to split graphs with convexity properties other than path, triad, star, comb, tree, and circular.  This would open up some new subclasses of split graphs having nice structural properties.
	% The graph problems that are NP-complete in split graphs, their complexity in convex split graphs is open. For the problems explored in convex split graphs, it is interesting to identify which convexity causes hardness? and for which convexity the problem is polynomial-time solvable?
	\bibliographystyle{unsrt}
	\bibliography{references}

\begin{thebibliography}{10}

\bibitem{white1985steiner}
Kevin White, Martin Farber, and William Pulleyblank.
\newblock Steiner trees, connected domination and strongly chordal graphs.
\newblock {\em Networks}, 15(1):109--124, 1985.

\bibitem{CHLEBIK20081264}
M.~Chlebík and J.~Chlebíková.
\newblock Approximation hardness of dominating set problems in bounded degree
  graphs.
\newblock {\em Information and Computation}, 206(11):1264--1275, 2008.

\bibitem{muller1987np}
Haiko M{\"u}ller and Andreas Brandst{\"a}dt.
\newblock The {NP}-completeness of {Steiner} tree and dominating set for
  chordal bipartite graphs.
\newblock {\em Theoretical Computer Science}, 53(2-3):257--265, 1987.

\bibitem{wald1983steiner}
Joseph~A Wald and Charles~J Colbourn.
\newblock {Steiner trees, partial 2--trees, and minimum IFI networks}.
\newblock {\em Networks}, 13(2):159--167, 1983.

\bibitem{wald1982steiner}
Joseph~A Wald and Charles~J Colbourn.
\newblock Steiner trees in outerplanar graphs.
\newblock In {\em Proc. 13th Southeastern Conf. on Combinatorics, Graph Theory,
  and Computing}, pages 15--22, 1982.

\bibitem{mohanapriya2021steiner}
A~Mohanapriya, P~Renjith, N~Sadagopan, et~al.
\newblock {Steiner Tree in $ k $-star Caterpillar Convex Bipartite Graphs--A
  Dichotomy}.
\newblock {\em arXiv preprint arXiv:2107.09382}, 2021.

\bibitem{chimani2012improved}
Markus Chimani, Petra Mutzel, and Bernd Zey.
\newblock {Improved Steiner tree algorithms for bounded treewidth}.
\newblock {\em Journal of Discrete Algorithms}, 16:67--78, 2012.

\bibitem{renjith2020steiner}
P~Renjith and N~Sadagopan.
\newblock The {Steiner tree in $K_{1, r}$-free split graphs—A Dichotomy}.
\newblock {\em Discrete Applied Mathematics}, 280:246--255, 2020.

\bibitem{damaschke1990domination}
Peter Damaschke, Haiko M{\"u}ller, and Dieter Kratsch.
\newblock Domination in convex and chordal bipartite graphs.
\newblock {\em Information Processing Letters}, 36(5):231--236, 1990.

\bibitem{panda2021dominating}
BS~Panda and Juhi Chaudhary.
\newblock Dominating induced matching in some subclasses of bipartite graphs.
\newblock {\em Theoretical Computer Science}, 885:104--115, 2021.

\bibitem{chen2016complexity}
Hao Chen, Zihan Lei, Tian Liu, Ziyang Tang, Chaoyi Wang, and Ke~Xu.
\newblock Complexity of domination, hamiltonicity and treewidth for tree convex
  bipartite graphs.
\newblock {\em Journal of Combinatorial Optimization}, 32(1):95--110, 2016.

\bibitem{pandey2019domination}
Arti Pandey and BS~Panda.
\newblock Domination in some subclasses of bipartite graphs.
\newblock {\em Discrete Applied Mathematics}, 252:51--66, 2019.

\bibitem{jiang2011tractable}
Wei Jiang, Tian Liu, and Ke~Xu.
\newblock Tractable feedback vertex sets in restricted bipartite graphs.
\newblock In {\em International Conference on Combinatorial Optimization and
  Applications}, pages 424--434. Springer, 2011.

\bibitem{jiang2011two}
Wei Jiang, Tian Liu, Tienan Ren, and Ke~Xu.
\newblock Two hardness results on feedback vertex sets.
\newblock In {\em Frontiers in Algorithmics and Algorithmic Aspects in
  Information and Management}, pages 233--243. Springer, 2011.

\bibitem{cormen2009introduction}
Thomas~H Cormen, Charles~E Leiserson, Ronald~L Rivest, and Clifford Stein.
\newblock {\em Introduction to algorithms}.
\newblock MIT press, 2009.

\bibitem{raman2008short}
Venkatesh Raman and Saket Saurabh.
\newblock {Short cycles make W-hard problems hard: FPT algorithms for W-hard
  problems in graphs with no short cycles}.
\newblock {\em Algorithmica}, 52(2):203--225, 2008.

\bibitem{dreyfus1971steiner}
Stuart~E Dreyfus and Robert~A Wagner.
\newblock The {Steiner} problem in graphs.
\newblock {\em Networks}, 1(3):195--207, 1971.

\bibitem{dom2009incompressibility}
Michael Dom, Daniel Lokshtanov, and Saket Saurabh.
\newblock Incompressibility through colors and ids.
\newblock In {\em International Colloquium on Automata, Languages, and
  Programming}, pages 378--389. Springer, 2009.

\bibitem{garey1979guide}
Michael~R Garey.
\newblock {\em {Computers and intractability: A Guide to the Theory of
  NP-Completeness}}.
\newblock WH Freeman \& Co, 1979.

\bibitem{ashok2015unique}
Pradeesha Ashok, Sudeshna Kolay, Neeldhara Misra, and Saket Saurabh.
\newblock Unique covering problems with geometric sets.
\newblock In {\em International Computing and Combinatorics Conference}, pages
  548--558. Springer, 2015.

\bibitem{ABUKHZAM2010524}
Faisal~N. Abu-Khzam.
\newblock A kernelization algorithm for d-hitting set.
\newblock {\em Journal of Computer and System Sciences}, 76(7):524--531, 2010.

\bibitem{vazirani2001approximation}
Vijay~V Vazirani.
\newblock {\em Approximation algorithms}, volume~1.
\newblock Springer, 2001.

\bibitem{BERTOSSI198437}
Alan~A. Bertossi.
\newblock Dominating sets for split and bipartite graphs.
\newblock {\em Information Processing Letters}, 19(1):37--40, 1984.

\end{thebibliography}
\end{document}